\documentclass[iop,apj,tighten,numberedappendix]{emulateapj}
\usepackage[breaklinks,colorlinks,urlcolor=blue,citecolor=blue,linkcolor=blue]{hyperref}

\usepackage{amsmath,amssymb}
\usepackage{cleveref}
\usepackage{graphicx}
\usepackage{bm}
\usepackage[toc,title,page]{appendix}
\usepackage{tocbibind}
\usepackage{xcolor,ulem}
\usepackage{multirow}

\newcommand{\Msun}{{\rm M}_\odot}

\usepackage{lineno}

\def\lsim{~\rlap{$<$}{\lower 1.0ex\hbox{$\sim$}}}
\def\gsim{~\rlap{$>$}{\lower 1.0ex\hbox{$\sim$}}}
 
\shorttitle{Electromagnetic Emission from Mergers in AGNs}
\shortauthors{Tagawa~et~al.}

\begin{document}
\title{Electromagnetic Flares from Compact-Object Mergers in AGN Disks:\\
Signatures and Predictions
}
\author{Hiromichi Tagawa\altaffilmark{1}, 
Zolt\'an Haiman\altaffilmark{2,3,4}, 
Shigeo S. Kimura\altaffilmark{5,6}, 
Hassen M. Yesuf\altaffilmark{1}, 
Hengxiao Guo\altaffilmark{1}
}
\affil{
\altaffilmark{1}Shanghai Astronomical Observatory, Shanghai, 200030, People$^{\prime}$s Republic of China\\
\altaffilmark{2}Department of Astronomy, Columbia University, 550 W. 120th St., New York, NY, 10027, USA\\
\altaffilmark{3}Department of Physics, Columbia University, 550 W. 120th St., New York, NY, 10027, USA\\
\altaffilmark{4}Institute of Science and Technology Austria, Am Campus 1, Klosterneuburg 3400, Austria\\
\altaffilmark{5}Astronomical Institute, Graduate School of Science, Tohoku University, Aoba, Sendai 980-8578, Japan\\
\altaffilmark{6}Frontier Research Institute for Interdisciplinary Sciences, Tohoku University, Sendai 980-8578, Japan\\
}
\email{E-mail: htagawa@shao.ac.cn}

\begin{abstract} 
Accretion disks in active galactic nuclei (AGN) are promising sites
for mergers of stellar-mass black holes (BHs) detectable via
gravitational waves (GWs). These environments facilitate both in-situ
formation and dynamical capture of compact objects, and their
subsequent mergers. The uncertain origin of GW events detected by
LIGO, Virgo and KAGRA motivates searching for accompanying
electromagnetic (EM) signatures.  
Here we investigate post-merger EM
flares associated with jets launched from merger remnants, as well as
from the shocked ambient gas as the jet
breaks out of the disk.  We find that jet breakout produces luminous
gamma-ray emission, detectable with MeV-band telescopes.  Cooling
emission from a shocked circum-BH minidisk, winds and background AGN-disk peaks in the UV and optical, with durations ranging from about an hour
to a month, and can be identified through year-long monitoring of
$\sim10^3$ AGNs with luminosities ranging from $\sim 10^{44}$
to $\sim 10^{45}~{\rm erg~s^{-1}}$.  With a single set of parameters,
this post-merger jet model produces gamma-ray, hard X-ray and optical flares similar to those claimed to be associated with GW events.
Furthermore, by incorporating a transition from a high- to
low-angular-momentum accretion state after the merger, the model
avoids excessive BH growth, alleviating tensions with hyper-Eddington
accretion scenarios.
\end{abstract}
\keywords{
transients 
-- stars: black holes 
--galaxies: active
}

\section{Introduction}

\label{sec:introduction}

To date, 
about 200 gravitational-wave (GW) events from mergers of stellar-mass black holes (BHs) and/or neutron stars (NSs) have been reported \citep{LIGO2025_O4a_population}. 
Although their astrophysical origins remain debated, active galactic nuclei (AGN) disks are a promising environment to facilitate BH mergers. 
During active phases, stellar-mass BHs can become embedded within AGN disks through interactions with the nuclear star cluster \citep{Ostriker1983,Bartos17,Wang2023_capture,Rowan2025b} or via in-situ star formation \citep{Levin2003,Stone17,Epstein-Martin2024,ChenY2023}. 
Subsequently, gas in AGN disks can facilitate binary formation \citep{Goldreich02,DeLaurentiis2023,Rowan2022,Dodici2024} and accelerate mergers \citep{Bartos17,McKernan17,Tagawa19,Xue2025}. 
Comparing observed BH masses \citep{Tagawa20_MassGap,Yang19a,Vaccaro2024,Xue2025,McKernan2025,Gayathri2021_AGN_O3,Gayathri2025,LiYinJie2025}, spins \citep{Yang19b,Tagawa20b_spin,Cook2025,Delfavero2025,Stegmann2025}, 
mass-spin correlations \citep{Tagawa2021_hierarchical,LiYinJie2024,LiYinJie2025_align}, 
eccentricities \citep{Samsing20,Tagawa20_ecc,Fabj2024,RomeroShaw2025}, 
center-of-mass acceleration \citep{Meiron17,Inayoshi17b,Han2024,Zwick2025,Tagawa2025}, 
lensing \citep{Leong2025,Samsing2025}
and spatial distribution relative to AGN \citep{Bartos17NatCom,Veronesi2022,Veronesi2023,Veronesi2024,Moncrieff2025,Zhu2025} 
can serve as tests for this channel. 
Significant uncertainties in this channel 
include the gas distribution surrounding BHs and merging binaries, 
particularly concerning the accretion processes onto BHs and the structure of AGN disks. 
These factors influence the demography and properties of mergers.

A distinctive prediction of the AGN-disk channel is the potential existence of electromagnetic (EM) counterparts accompanying BH mergers~\citep{Bartos17,Stone17,Ford2025}. 
Several candidate counterparts have been reported, including optical \citep{Graham20,Graham2023}, gamma-ray \citep{Connaughton2016,Bagoly2016}, and X-ray flares \citep{2024GCN.38308....1D,2024GCN.38345....1W}. However, their associations with GW events are still under debate \citep{Ashton2020,Palmese2021,Veronesi2024_flare}, 
with interpretations of optical counterparts depending on 
criteria such as light-curve shape or magnitude variations \citep{Graham20,Palmese2021,He2025}.

Various models have been proposed for EM flares generated by post-merger BHs \citep[e.g.,][]{Perna2016,Loeb2016,deMink2017,McKernan2019_EM}. 
Promising scenarios involve shocks driven by jets launched from the post-merger BH. 
The radiation mechanisms include 
thermal \citep{Tagawa2023_SC,Chen2024}, non-thermal \citep{Tagawa2023}, and free-free or bound-free emission \citep{RodriguezRamirez2023}. 
Two representative pathways for EM emission following mergers are: 
(i) jet reorientation, allowing the jet to interact with fresh AGN gas \citep{Tagawa2023}, 
and (ii) gas capture following recoil kicks that power jets and flares \citep{deMink2017,Graham20,Chen2024}.

A key theoretical challenge is producing high accretion rates and strong jets 
without causing excessive BH growth. 
In practice, these episodes must therefore be very brief and associated with the merger. 
Prior studies of EM emission from merging BHs in AGN disks often assume sustained hyper-Eddington accretion 
\citep[e.g.,][]{Graham20,Tagawa2023, Tagawa2023_solitary,McPike2026}, which can lead to rapid 
BH overgrowth. 
In such a situation, where the captured gas accretes onto BHs without regulation, Soltan's argument \citep{Yu2002} could be violated, by depleting most of the inflow gas through accretion onto the stellar-mass BHs, while the central supermassive BHs (SMBHs) grow primarily through mergers with surrounding BHs rather than through gas accretion \citep{Tagawa2022_BHFeedback}. Additionally, this could challenge the observed lack of intermediate-mass BHs in the vicinity of the Galactic center \citep{Gravity2024}, as they could form during previous AGN episodes \citep{Su2010} if efficient growth occurs \citep{Tagawa2022_BHFeedback}.

To address this, 
we consider a post-merger transition from 
an adiabatic inflow-outflow solution (ADIOS)-like 
high-angular-momentum accretion mode \citep{Blandford1999} to a low-angular-momentum, zero-Bernoulli accretion (ZEBRA) flow \citep{Coughlin2014}. 
In the ADIOS state, radiation pressure limits accretion by diffusing photons with the help of magnetic fields, which eject much of the inflowing gas. 
Conversely, in the ZEBRA state, low-angular-momentum gas circularizes to form 
a quasi-spherical disk (see \S~\ref{sec:prob_gamma}), 
where 
efficient advection relative to photon diffusion suppresses winds, enabling hyper-Eddington accretion and strengthening possible Blandford–Znajek jets \citep{Blandford1977}. 
The transition between these states depends on the circularization radius ($r_{\rm circ}$) relative to the photon trapping radius ($r_{\rm trap}$). If $r_{\rm circ} > r_{\rm trap}$, the system remains in the ADIOS regime; if $r_{\rm circ}< r_{\rm trap}$, the system operates in the ZEBRA-like advection-dominated regime \citep{Begelman2017}. 
This framework is supported by 
3D general relativistic radiation magnetohydrodynamic simulations \citep{Sadowski2016,Fragile2025} and observations of X-ray binaries and tidal disruption events \citep{Poutanen2007,Zauderer2011,Begelman2017}.

Incorporating this transition, we predict the properties of EM flares emanating from merging BHs. 
To construct a comprehensive model, 
we consider emission from multiple components including winds and 
a circumbinary or circum-remnant disk (CBD), in addition to the AGN disk gas investigated in \citet{Tagawa2023,Tagawa2023_SC}. 
Additionally, we extend this model to describe analogous emission resulting from 
shocked gas produced by supernova (SN) explosions occurring within AGN disks. 
We also explore the observational prospects for detecting these flares (\S~\ref{section:observability}), 
and discuss how to constrain the gas distribution surrounding merging BHs based on EM counterparts (\S~\ref{sec:cooling_dependence}, \S~\ref{sec:results_optical}).

As the gas distribution, 
we adopt two widely used prescriptions for AGN disks. 
One is the SG model \citep[][]{Sirko03} characterized by the typical $\alpha$ viscosity, 
while the other is the TQM model \citep{Thompson05}, which features more efficient angular momentum transfer. 
While the SG model is not self-consistent 
in the outer regions due to a lack of gas depletion by star formation 
(and it also violates mass constraints there, \citealt{Hopkins2026}), 
it provides a useful basis for comparing the two models to understand the influence of angular momentum transfer efficiency and corresponding disk density. 
These models yield distinct emission signatures, which can help observationally constrain the properties of AGN disks.

This paper primarily focuses on thermal emission from jets and SN explosions, as these are inevitably produced following strong shocks and tend to be brighter. 
We discuss how to distinguish these flares from other transients and typical AGN variability, compare our predictions with reported optical and gamma-ray candidate counterparts to LVK events, and explore how observed flare properties can constrain AGN disk models and the properties of shocked components.

This paper is organized as follows. In \S~\ref{sec:overview}, 
we provide an overview of our model; 
\S~\ref{sec:method} details the EM-flare modeling approach; 
\S~\ref{sec:results} presents our main findings, 
\S~\ref{sec:discussion} discusses the broader implications of our results, 
and 
\S~\ref{sec:conclusions} summarizes our key conclusions.

\begin{figure*}
\begin{center}
\includegraphics[width=180mm]{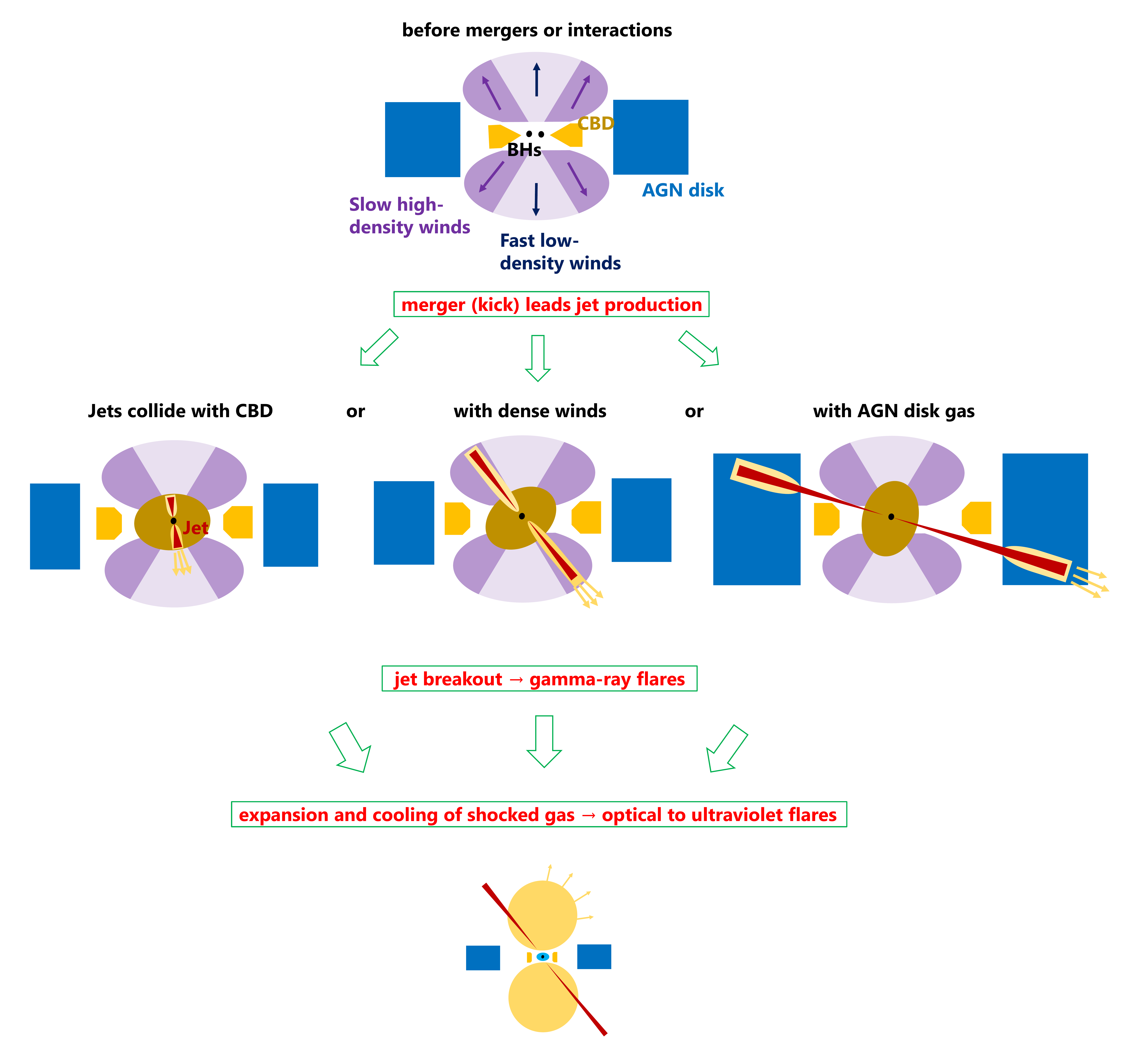}
\caption{
A schematic illustration of EM flares
associated with compact-object mergers within an AGN disk. 
{\bf Top row:} During hyper-Eddington accretion of 
high-angular-momentum gas onto the merging binary, strong outflows are driven (ADIOS mode). If a persistent jet is present, it propagates through a cavity in the AGN disk without significantly interacting with dense material.
{\bf Middle row:} After the merger, the accretion mode transitions to a low-angular-momentum flow (ZEBRA mode), resulting in the formation of 
a quasi-spherical disk 
and a strong jet. 
The jet then collides with 
the CBD whose inner region now manifests as a ZEBRA 
flow (left), with
dense winds (middle), or with AGN disk gas (right). 
These interactions drive shocks, producing bright shock breakout flares.
{\bf Bottom row:} The hot, shocked gas expands and cools outside the AGN disk, giving rise to an additional flare. 
}
\label{fig:schematic}
\end{center}
\end{figure*}

\section{Methods}

\label{sec:method}

We outline our model for accretion onto BHs and the associated EM emission resulting from compact-object mergers within AGN disks. 
Fig.~\ref{fig:schematic} provides a schematic overview of the key processes: accretion mode changes, jet launching, and emission following shock breakout.

\subsection{Overview of the model}

\label{sec:overview}

We present predictions for observable EM counterparts to compact-object mergers occurring within AGN disks. 
Our analysis considers emission from several components across a broad parameter space, based on two key assumptions. 
First, we assume that accretion-state transitions depend on the ratio of the circularization radius $r_{\rm circ}$ to the photon-trapping radius $r_{\rm trap}$. 
ADIOS prevents excessive BH growth by regulating accretion (Appendix~\ref{sec:growth}), 
while ZEBRA flow enables bright emission associated with BH mergers (\S~\ref{sec:model_accretion}).

The basic emission processes operate as follows: 
Before mergers, 
stellar-mass BH binaries embedded in AGN disks 
capture gas and 
launch dense winds \citep[e.g.,][]{Poutanen2007} that carve cavities within the AGN disk. 
When a BH receives a recoil kick during a merger or after binary–single interactions, 
strong shocks develop within the CBD. These shocks can temporarily modify the CBD's angular-momentum orientation. If the binary's orbital angular momentum before the merger is misaligned with the primary BH's spin, the spin direction may also change \citep{Tagawa20b_spin}. 
Note that the jet strength is significantly enhanced after merger due to the enhancement of BH spins at merger \citep{Buonanno08}, possible transition of accretion mode (\S\ref{sec:model_accretion}), and enhancement of accretion by shocks (Appendix~B of \citealt{Tagawa2023}). 
Since jets are launched along the BH spin axis and tend to align with the CBD's angular momentum \citep{Liska2018,Polko2017}, such reorientations can cause the jet to change direction post-merger. A reoriented jet collides with the disk gas, winds, and the CBD, generating strong shocks. Once photon diffusion overtakes shock propagation, photons escape, leading to breakout emission \citep[e.g.,][]{Nakar2010}. Subsequent diffusion from deeper layers gives rise to shock-cooling emission \citep[e.g.,][]{Arnett1980}.
We expect both thermal and non-thermal emission in the breakout phase, 
as described by \citet{Tagawa2023} and \citet{Tagawa2023_solitary}, provided that collisionless shocks form (but see \S~\ref{sec:nonthermal}). On the other hand, 
during the cooling phase, thermal emission is likely to dominate \citep{Tagawa2023_SC}, similar to  supernova emission \citep{Arnett1980}.

\subsection{Accretion, kicks, and jet production}

\label{sec:model_accretion}

We begin by describing the two different accretion modes we adopted for stellar-mass BHs embedded in an AGN disk. 
We estimate the gas capture rate by a BH (${\dot M}_{\rm cap}$) using a modified Bondi–Hoyle–Lyttleton formula (Eq.~1 of \citealt{Tagawa2022_BHFeedback}). 
The formation of gaps--reducing the local disk surface density--is incorporated following the prescription of \citet{Kanagawa18} as implemented in \citet{Tagawa19}, where the gap depth is determined by the balance between viscous angular momentum flux and the gravitational torque exerted by the BH. 
Following \citet{Begelman2017}, 
we assume an ADIOS-like inflow–outflow solution for $r_{\rm circ}>r_{\rm trap}$, and a ZEBRA-like, advection-dominated flow for  $r_{\rm circ}<r_{\rm trap}$ (\S~\ref{sec:introduction}).

When the BH is kicked, 
the values of the circularization radius $r_{\rm circ}$ and trapping radius $r_{\rm trap}$ evolve. 
Prior to the kick, 
the gas captured by the BH from the AGN disk circularizes at 
\begin{eqnarray}
\label{eq:r_circ}
r_{\rm circ,bk} &= &f_{\rm circ}r_{\rm Hill}\nonumber\\
&\sim& 2\times 10^{15}
\left(\frac{R_{\rm BH}}{0.1~{\rm pc}}\right)
\left(\frac{m_{\rm BH}/M_{\rm SMBH}}{10^{-6}}\right)^{1/3}\nonumber\\
&&\left(\frac{f_{\rm circ}}{1}\right)~{\rm cm}, 
\end{eqnarray}
where $R_{\rm BH}$ is the distance of the BH from the SMBH, $r_{\rm Hill}=R_{\rm BH}(m_{\rm BH}/3M_{\rm SMBH})^{1/3}$ is the Hill radius, 
$m_{\rm BH}$ and $M_{\rm SMBH}$ are the masses of the BH and SMBH, respectively, and 
$f_{\rm circ}$ is a factor representing the fraction of the circularization radius over the Hill radius.  
The factor $f_{\rm circ}$ depends on the pressure-gravity balance and is estimated as \citep{Sagynbayeva2024}
\begin{align}
\label{eq:f_circ}
f_{\rm circ}\sim {\rm min}\{1,0.01(m_{\rm BH}/M_{\rm SMBH})^2(R_{\rm BH}/H_{\rm AGN})^6\}, 
\end{align}
where $H_{\rm AGN}$ is the scale height of the AGN disk. 
Pre- and post-kick quantities are distinguished by subscripts $"{\rm bk}"$ and $"{\rm ak}"$ respectively. 
The trapping radius before the kick is given by \citep[e.g.][]{Kato2008}
\begin{eqnarray}
\label{eq:r_trap}
r_{\rm trap,bk}&\sim &{\dot m}_{\rm cap,bk} r_{\rm g} \nonumber\\
&\sim& 3\times 10^{13}
\left(\frac{{\dot m}_{\rm cap,bk}}{10^6}\right)
\left(\frac{m_{\rm BH}}{100~\Msun}\right)~{\rm cm}, 
\end{eqnarray}
where $r_{\rm g}=Gm_{\rm BH}/c^2$ is the gravitational radius of the BH, $G$ is the gravitational constant, $c$ is the speed of light, 
${\dot m}_{\rm cap}={\dot M}_{\rm cap}/{\dot M}_{\rm Edd}\eta_{\rm rad}$ is the Eddington ratio for the capture rate, 
${\dot M}_{\rm cap}$ is the gas capture rate, 
${\dot M}_{\rm Edd}=L_{\rm Edd}/\eta_{\rm rad}c^2$ is the Eddington rate, 
$L_{\rm Edd}$ is the Eddington luminosity, and 
$\eta_{\rm rad}$ is the radiative efficiency set to $\eta_{\rm rad}=0.1$.

Prior to the merger, the condition $r_{\rm circ,bk}>r_{\rm trap,bk}$ implies that accretion is suppressed in the ADIOS mode, 
which occurs if  
\begin{align}
\label{eq:adios_con}
{\dot m}_{\rm cap,bk}\lesssim 6\times 10^{7}
\left(\frac{m_{\rm BH}}{100~\Msun}\right)^{-1}
\left(\frac{r_{\rm circ,bk}}{2\times 10^{15}~{\rm cm}}\right). 
\end{align}

Immediately after the merger or a binary-single interaction, 
the BH receives a recoil kick, generating shocks in the CBD. 
These shocks develop between radii $\sim r_{\rm b}$ and $\sim r_{\rm ub}$, within which a significant fraction ($\sim 20$--$50\%$) of the gas becomes circularized on a dynamical timescale \citep{Rossi2010}, 
where 
\begin{align}
\label{eq:rb}
r_{\rm b}=f_{\rm b} r_{\rm kick}
\end{align} and 
\begin{align}
\label{eq:rub}
r_{\rm ub}=f_{\rm ub} r_{\rm kick}
\end{align} 
are the radii inside and outside which all gas is bound and unbound to the kicked BH, respectively, 
\begin{align}
\label{eq:rkick}
r_{\rm kick}=\frac{Gm_{\rm BH}}{v_{\rm kick}^2}
\sim 2\times 10^{13}\left(\frac{m_{\rm BH}}{50~\Msun}\right)\left(\frac{v_{\rm kick}}{200~{\rm km/s}}\right)^{-2}~{\rm cm} 
\end{align} 
is a characteristic radius at a kick, and constants $f_{\rm b}$ and $f_{\rm ub}$ depend on the kick direction, 
with typical values of $\sim 0.2$ and $\sim 5$, respectively, for kick angles with respect to the CBD plane being $\theta_{\rm kick}\sim 0$--$40^\circ$ \citep{Rossi2010}. 
Since shocked gas typically circularizes near $r_{\rm b}$, 
we set $r_{\rm circ,ak}=r_{\rm b}$. 
Additionally, 
shocks caused by kicks can also temporarily enhance the inflow rate by up to a factor $f_{\rm inc} \sim 60$ \citep{Tagawa2023}, affecting the trapping radius.

As a result, the post-kick bound gas may satisfy $r_{\rm circ,ak}< r_{\rm trap,ak}$, 
establishing a ZEBRA-like accretion state if 
\begin{align}
\label{eq:zebra_con}
{\dot m}_{\rm inf,ak}(r_{\rm b}) \gtrsim 2\times 10^{5}\left(\frac{v_{\rm kick}}{300~{\rm km~s^{-1}}}\right)^{-2}\left(\frac{f_{\rm b}}{0.2}\right),
\end{align}
where ${\dot m}_{\rm inf}(r)={\dot M}_{\rm inf}(r) c^2\eta_{\rm rad}/L_{\rm Edd}$ is the Eddington ratio of the inflow rate  (${\dot M}_{\rm inf}(r)$) at a distance $r$ from the stellar-mass BH.

Before the kick, 
accretion typically proceeds via the ADIOS mode, 
with the inflow rate decreasing inwards as  
\begin{align}
\label{eq:min_adios}
{\dot M}_{\rm inf,bk}(r)={\dot M}_{\rm cap,bk}\left(\frac{r}{r_{\rm trap,bk}}\right)^p, 
\end{align} 
where $p$ is the power-law slope. 
We adopt $p=1$ following the recent study on 3D general-relativistic radiation magnetohydrodynamic simulations (\citealt{Fragile2025}), which suggest a scenario that may alleviate the overgrowth problem. 
However, radiation-hydrodynamical simulations tend to imply lower values of $p$ (e.g., \citealt[][]{Hu2022,Toyouchi2024}), in which the overgrowth problem persists. 
Sensitivity to $p$ is explored in Fig.~\ref{fig:flare_dep2} of Appendix~\ref{sec:parameter_dependence}.

The post-kick accretion rate onto the BH is modeled as 
\begin{align}
\label{eq:macc_af}
{\dot M}_{\rm acc,ak}=f_{\rm inc}{\dot M}_{\rm cap,bk}  {\rm min}\left\{1, \left(\frac{r_{\rm kick}}{r_{\rm trap,bk}}\right)^p \right\},
\end{align}
which accounts for the enhancement due to shocks and the suppression associated with the ADIOS mode when $r_{\rm kick}$ is within $r_{\rm trap,bk}$  (Eq.~\ref{eq:min_adios}) 
\footnote{
In Eq.~\eqref{eq:macc_af}, $r_{\rm kick}$ is substituted for $r$ in Eq.~\eqref{eq:min_adios}, representing 
the region where most of the bound gas initially resides. 
}$^,$
\footnote{
After re-entering the AGN disk, the BH can accrete additional gas via the Bondi-Hoyle-Lyttleton mechanism; this process is not modeled here. 
}.

If the condition $r_{\rm b}<r_{\rm circ,bk}$ is satisfied, strong shocks develop in the CBD following kicks. 
The shocks circularize at smaller radii, which can also increase $r_{\rm trap,ak}$ and potentially trigger a transition to the ZEBRA state. After this transition the accretion rate is greatly enhanced, 
leading to the formation of powerful jets. 
This high-accretion, jet-producing state is established after the dynamical time at $r_{\rm b}$ 
and persists for the viscous timescale there.

For shock-cooling emission, since the photon diffusion timescale exceeds the dynamical timescale, 
we neglect any additional GW-EM delays in the accretion-rate enhancement; 
thus, the GW-EM delay is predominantly governed by the diffusion timescale. 
Additionally, because 
the jet breakout timescale is typically shorter than the viscous timescale ($t_{\rm vis}$, Eq.~\ref{eq:t_vis}), 
we assume the jet's kinetic power (Eq.~\ref{eq:lj_macc} below) remains constant until breakout. 
This assumption is valid if the component size is less than $\sim 10^{15}~{\rm cm}(t_{\rm vis}/3\times 10^5~{\rm s})(v_{\rm h}/0.1c)$ (Figs.~\ref{fig:sizes_tqm} and \ref{fig:sizes_sg}), where $v_{\rm h}$ is the jet head velocity in the component. If this condition is not satisfied, the breakout emission is unobservable, and the shock-cooling emission becomes dimmer (see Eq.~13 of \citealt{McPike2026}). 
We discuss the effects of these timescales on shock-breakout emission in \S~\ref{section:delay_time}. 
Gas located outside $r_{\rm ub}$ is unbound and is assumed to 
eventually rejoin the background AGN disk.

\subsection{Model components}

Due to the rapid accretion described in the previous section, 
jets are expected to be launched and consequently collide with nearby gas,
producing shocks and bright EM emission. 
In the following, we describe the properties of the nearby gas contributing to this emission, including several distinct components.

\subsubsection{Jet}

\label{section:jet}

When the condition $r_{\rm circ}<r_{\rm trap}$ is satisfied 
and the BHs possess significant spin, 
we assume that Blandford–Znajek jets are launched with the power given by
\begin{align}
\label{eq:lj_macc}
L_{\rm j}= \eta_{\rm j}{\dot M}_{\rm acc} c^2,
\end{align}
provided that a CBD is in a state of a magnetically arrested disk, achieved through the accumulation of strong magnetic fields within AGN disks (Appendix A.1 of \citealt{Tagawa2022_BHFeedback}), where 
$\eta_{\rm j}$ is the jet conversion efficiency. 
The accretion rates onto the BHs remain steady over longer timescales compared to the breakout timescale of the jet from surrounding components, enabling a constant jet power (Eq.~\ref{eq:lj_macc}). Hence, the time evolution of magnetic fields and accretion rates is not analogous to that in gamma-ray bursts \citep[e.g.,][]{Perna2021_GRBs,Gottlieb2023}.

Due to wind and jet production, a cavity forms in the AGN disk around the BH (\S~\ref{sec:agn_disk}). 
Although the cavity forms, 
accretion can be intermittent, with an active duty cycle $\gtrsim 10\%$ \citep{Tagawa2022_BHFeedback}. A BH can therefore continue to accrete during periods of depletion of gas capture.

\subsubsection{Wind}

Given the high gas densities in AGN disks, 
BH gas-capture rates can far exceed the Eddington accretion rate. 
Under such circumstances, 
the advection of gas is more efficient compared to 
the diffusion of photons within the trapping radius ($r_{\rm trap}$). 
Both within and slightly outside 
$r_{\rm trap}$, 
radiation pressure exceeds gravitational forces, 
causing the CBD to become geometrically thick. 
A substantial fraction of the gas can be ejected in the ADIOS state, 
regulating the BH mass accretion rate to near the Eddington accretion rate. 
Here, we assume the wind production rate is approximately equal to the capture rate, 
\begin{align}
\label{eq:md_wind}
{\dot M}_{\rm wind} \approx {\dot M}_{\rm cap} \gg {\dot M}_{\rm acc,bk}.
\end{align}
Since a significant portion of the winds are launched at the trapping radius ($r_{\rm trap}$), 
we assume the wind velocity is given by \citep[e.g.][]{Poutanen2007}
\begin{align}
\label{eq:v_wind}
v_{\rm wind}\simeq \left(\frac{Gm_{\rm BH}}{r_{\rm trap}}\right)^{1/2}
= c {\dot m}_{\rm cap}^{-1/2}. 
\end{align}

Considering the gravitational force exerted by the SMBH, 
the winds can reach a maximum height of $H_{\rm \Omega} = v_{\rm wind}/\Omega_{\rm SMBH}$ above and below the BH, 
where 
$\Omega_{\rm SMBH}$ is the orbital angular velocity at the BH's location around the SMBH. 
This limitation appears as long as the local Keplerian velocity about the SMBH exceeds the wind velocity. 
Additionally, we assume that the shock breakout and cooling emission are produced 
beyond a distance from the BH 
where photons propagate faster than the wind expansion speed. 
At this distance ($r=H_{\tau}$), 
\begin{align}
\label{eq:h_tau}
v_{\rm ej}=\frac{c}{H_{\tau}\kappa \rho_{\rm wind}(H_{\tau})}
\end{align}
is satisfied, 
where  
$\rho_{\rm wind}(r)$ is the wind density at the distance $r$ from the BH, 
$\kappa$ is the opacity, assumed to be $0.4~{\rm cm^2~g^{-1}}$, 
and $v_{\rm ej}$ is the expansion velocity of the shocked gas. 
We set the wind size to $H_{\rm wind}={\rm min}(H_{\Omega},H_\tau)$.

The wind density at $r$ can be approximated as 
\begin{align}
\label{eq:rho_wind}
{\dot M}_{\rm wind}= S_{\rm wind}(r)\rho_{\rm wind}(r) v_{\rm wind}, 
\end{align}
where 
$S_{\rm wind}(r)=\Omega_{\rm wind} r^2$ is the surface area at $r$ and 
$\Omega_{\rm wind}$ is the solid angle 
into which winds are launched. 
\citet{Poutanen2007} 
inferred $\Omega_{\rm wind}\sim (0.5$--$0.85)4\pi$, 
from observations of ultraluminous X-ray sources. 
We adopt $\Omega_{\rm wind} \sim 0.7\times 4\pi$, 
corresponding to dense winds launched within $\pm \pi/4$ of the CBD plane. 
For simplicity, we neglect the fast, low-density winds expected in the polar direction of the disk (Fig.~\ref{fig:schematic}).

\subsubsection{Circumbinary disk (CBD)}

\label{sec:cbd}

Prior to mergers or binary-single interactions, gas is captured by the BHs at hyper-Eddington rates, 
forming a geometrically thick accretion disk with a size of $\sim {\rm min}(r_{\rm trap,bk},r_{\rm circ,bk})$. 
If $r_{\rm trap,bk}<r_{\rm circ,bk}$, the BH is further surrounded by 
a geometrically thin accretion disk within a radius comparable to the circularization radius ($r_{\rm circ,bk}$).

When a BH experiences a recoil kick during a merger or binary-single interaction, 
shocks develop within the CBD, and the accretion state can change (see $\S~\ref{sec:model_accretion}$). 
Specifically, 
the size of the CBD becomes limited to $r_{\rm CBD} = r_{\rm circ,ak}={\rm min}(r_{\rm b},r_{\rm circ,bk})$, 
which can be reduced by a factor of $\gtrsim 100$ (Eqs.~\ref{eq:r_circ}, \ref{eq:rb}, and \ref{eq:rkick}) in our fiducial model.

Within the trapping radius, 
we assume the scale height-to-radius ratio of the CBD to be $h_{\rm CBD}\sim 1$ \citep{Kato2008}. 
The surface density at radius $r$ is approximated as 
\begin{align}
\label{eq:sigma_cbd}
\Sigma_{\rm CBD} (r)=
\frac{{\dot M}_{\rm inf}(r)}
{3 \pi \alpha r^2  h_{\rm CBD}(r)^2 \Omega_{\rm BH}(r)}
\end{align}
(e.g., SG), 
where $\Omega_{\rm BH}(r)$ is the local orbital angular velocity at $r$ from the BH,  
and $\alpha$ is the Shakura-Sunyaev viscosity parameter.

\subsubsection{AGN disk} 

\label{sec:agn_disk}

We assume that initially, compact objects and stars are embedded within the AGN disk. 
A cavity forms around these compact objects in the AGN disk due to wind ejection. 
This occurs because, at the scale height of the AGN disk, the ram pressure exerted by the wind exceeds the combined radiation and gas pressures of the AGN disk gas 
provided that the wind velocity significantly surpasses the disk's sound speed, $c_{\rm s}$. 
This condition can be roughly expressed as $v_{\rm wind}v_{\rm sh}^2\gtrsim c_{\rm s}^3$, where $v_{\rm sh}$ is the shear velocity at the Hill radius.

This wind-driven clearing results in a 
cavity with a width of a few times $H_{\rm AGN}$ \citep{Kompaneets1960,Kimura2021_BubblesBHMs,Tagawa2022_BHFeedback}. 
Consequently, jets generally escape the disk unless they are strongly inclined. For inclined jets, we approximate the distance to the AGN surface along the jet's path as $f_{\rm corr} H_{\rm AGN}$, 
where we set $f_{\rm corr}=2$, as a typical value. 

In addition to the cavity, a gap can form from gravitational torques exerted by the BH, whose width is typically much wider than the size of the cavity \citep{Kanagawa2016}, and its surface density remains nonzero. 
Considering gap formation, we reduce the surface density of the AGN disk following \citet{Kanagawa18}. For simplicity, we keep the temperature and scale height fixed.

The emission produced when the jet interacts with the AGN disk gas has been discussed in \citet{Tagawa2023} and \citet{Tagawa2023_SC}. 
We revisit this process in subsequent sections to analyze its parameter dependence and compare it with emission arising from other shocked gaseous components.

\subsection{Shock formation and evolution}

\label{section:collisions}

When the jet collides with surrounding components -- such as winds, CBDs, or AGN disk gas -- shocks form and propagate. 
Initially, the jet direction aligns with the spin axis of the merged remnant, but it gradually transitions to align with the angular momentum direction of the CBD \citep{Liska2018,Polko2017}. 
Depending on the relative orientations of the jet and the planes of the CBD and the AGN disk, the jet may first collide with the CBD, winds, or the AGN disk (Fig.~\ref{fig:schematic}). It is also possible for the jet to collide with more than one of these components.

If a jet encounters multiple components, emission from an inner component can be absorbed or scattered by outer layers. 
Typically, the observed signal is 
dominated by the outermost radiating component, except for high-energy photons that can escape more readily.

Once the jet interacts with winds, CBDs, or the AGN disk gas, two shocks develop: a forward shock propagating into the surrounding medium and a reverse shock traveling back into the jet. The region between these shocks is called the jet head \citep[e.g.][]{Matzner2003}. Surrounding the jet is shocked material known as the cocoon. 
The dynamical evolution of the jet is governed by the interaction between the jet and the cocoon.

We employ the formulae provided in \citet{Tagawa2023} 
to compute shock velocities and breakout/diffusion timescales in both the non-relativistic and relativistic regimes. 

\subsection{Emission processes}

Shock breakout occurs when photon diffusion 
becomes more rapid than 
shock propagation, producing a brief breakout emission \citep{Rabinak2011,Nakar2010,Nakar2012}. 
Subsequent photon diffusion from deeper layers yields a longer-lived shock-cooling light curve \citep{Arnett1980,Sapir2017,Morag23}. We calculate breakout luminosities and times using the formulae from \citet{Tagawa2023}, and model the cooling emission following \citet{Tagawa2023_SC}. 
In non-relativistic breakout regimes, 
light-travel time smearing is negligible, 
whereas it is accounted for in relativistic cases.

\begin{table*}
\begin{center}
\caption{
The table lists model parameters (columns 1 and 2), fiducial values (col.~3), ranges (col.~4), the introductory section (col.5), and references (col.6).
}
\label{table:parameter_fiducial}
\hspace{-5mm}
\begin{tabular}{p{1cm}|p{4.6cm}|p{1.5cm}|p{2.4cm}|p{1.2cm}|p{4cm}}
\hline 
Symbol&Parameter & Fiducial&Range&Section& References \\
\hline\hline
$m_{\rm BH}$&Mass of the merger remnant 
& $50\,{\Msun}$& $\sim 10$--$200~{\Msun}$&$\S~\ref{sec:model_accretion}$&\citet{Abbott21_GWTC3}\\\hline
$p$&Power-law index for the reduction of the inflow rate due to wind losses & $1$&$\sim 0$--$1$&$\S~\ref{sec:model_accretion}$&\citet{Kitaki2021,Hu2022,Fragile2025}\\\hline
$\eta_{\rm rad}$&Conversion of efficiency of rest-mass energy to radiation& $0.1$& $\sim 0.03$--$ 0.7$&$\S~\ref{sec:model_accretion}$&\citet{Shakura73,Fragile2025}\\\hline
$f_{\rm corr}$&Jet propagation distance to the AGN surface over the AGN scale height & $2$&$\gtrsim 1$&$\S~\ref{sec:model_accretion}$&\citet{Rossi2010,Tagawa2023}
\\\hline
$f_{\rm inc}$&Post-kick increase in the accretion rate due to shocks in the CBD & $10$&$\sim 1$--$60$&$\S~\ref{sec:model_accretion}$&\citet{Rossi2010,Tagawa2023}
\\\hline
$f_{\rm b}$ ($f_{\rm ub}$)&Dimensionless radius inside (outside) which all gas remains bound (unbound) to the kicked BH & $0.2$ ($5$)&
$\sim 0.17$--$1$ \par ($1$--$5.8$)
&$\S~\ref{sec:model_accretion}$ \, (Eqs.~\ref{eq:rb},~\ref{eq:rub}) &\citet{Rossi2010}
\\\hline
$v_{\rm kick}$&Recoil kick velocity after BH merger & $200~{\rm km~s^{-1}}$&$\sim 0$--$4000~{\rm km~s^{-1}}$&$\S~\ref{sec:model_accretion}$&\citet{Schnittman2007,Campanelli2007}
\\\hline
$\eta_{\rm j}$&Energy conversion efficiency to jet& $0.5$& $\sim 0.1$--$ 0.8$&$\S~\ref{section:jet}$&\citet{Tchekhovskoy2011,Narayan2021,Abbott21_GWTC3}\\\hline
$\alpha$&Viscous parameter for inner regions of AGN disks and CBDs& $0.1$&$\sim 0.01$--$0.3$&$\S~\ref{sec:cbd}$&\citet{King07,Jiang+2014}\\\hline
$\theta_{\rm 0}$&Opening angle of the injected jet & $0.2~{\rm rad}$&$\sim 0.01$--$0.3~{\rm rad}$&$\S~\ref{sec:parameters}$&\citet{Berger2014,Hada2018,Hada2019}\\\hline
$m$&Angular momentum transfer parameter due to global torques& $0.2$&$\lesssim 1$&$\S~\ref{sec:parameters}$&\citet{Thompson05,Collin2008}\\\hline
\end{tabular}
\end{center}
\end{table*}

\subsection{Model parameters}

\label{sec:parameters}

This section outlines the fiducial values for the model parameters. 
Table~\ref{table:parameter_fiducial} lists these values, their possible ranges, the sections of this paper where they are introduced, 
and relevant references. 
The fiducial values are set as follows:
\begin{itemize} \item
Merged remnant mass: $50~{\Msun}$ 
\item
Power-law index for the reduction of the inflow rate: $p=1$ 
\item
Energy conversion efficiency to radiation: $\eta_{\rm rad}=0.1$ 
\item
Correction factor for distance to AGN gas 
$f_{\rm corr}$: 2 (\S~\ref{sec:agn_disk}) 
\item
Post-kick increase in accretion rate due to shocks in the CBD: $f_{\rm inc}=10$ 
\item
Bound radius factor: $f_{\rm b}=0.2$, where $r_{\rm b}=f_{\rm b}r_{\rm kick}$
\item
Recoil kick velocity after BH merger: $200~{\rm km~s^{-1}}$ 
\item
Energy conversion efficiency to jet: $\eta_{\rm j}=0.5$ \citep{Tchekhovskoy2011,Narayan2021}, considering high spin magnitudes for merger remnants \citep{Buonanno08} 
\item
Thin disk viscosity parameter: $\alpha=0.1$
\item
Opening angle of the injected jet: $\theta_0 = 0.2~{\rm rad}$ (note that a smaller value is adopted in \S~\ref{sec:discussion}) 
\end{itemize}

For the AGN disk model, we employ two typical disk models proposed by SG and TQM. 
We utilize the default setup of the AGN-disk modeling tool \textit{pAGN} \citep{Gangardt2024}, which calculates the density, scale height, temperature, and inflow rate as a function of the distance from the SMBH for both models. 
In the TQM model, the inflow velocity is set to a factor $m$ times the local sound speed, with $m=0.2$. 
This prescription implies that angular momentum transfer is approximately the disk's aspect ratio times more efficient than in the $\alpha$-disk, resulting in a lower gas density by this factor.

\subsection{Shocks caused by supernova explosions} 

\label{section:explosions}

Although our main focus is on BH-merger-driven jets, the same shock and radiative framework applies to explosive events in AGN disks, such as core-collapse SNe, NS or white dwarf mergers. 
For illustrative SN calculations, assuming type~Ia supernovae, 
we adopt: 
\begin{itemize} \item 
Explosion energy: $E_{\rm SN}=10^{51}~{\rm erg}$
\item 
Ejecta mass: $m_{\rm SN}=1~{\Msun}$ 
\end{itemize}
The ejecta expand spherically, with 
an initial velocity: $v_{\rm SN}=(2E_{\rm SN}/m_{\rm SN})^{1/2}\simeq 10^9~{\rm cm~s^{-1}}$. 
After sweeping up a comparable mass of gas from the AGN disk $m_{\rm SN}$, 
the ejecta decelerate adiabatically. 
To calculate the gap structures surrounding the progenitor star, we set the progenitor mass to $1~{\Msun}$.

Massive stars embedded in an AGN disk may create cavities through their strong stellar winds. 
We find that such cavities do not form in the SG model.  
In the TQM model, however, a cavity forms outside a radius: 
$R_{\rm BH}\gtrsim 10^6~r_{\rm g} ({\dot M}_{\rm SMBH}/{\dot M}_{\rm Edd})$, 
assuming winds are emitted from a progenitor star with a velocity of $10^3~{\rm km~s^{-1}}$ 
and a mass-loss rate of $10^{-5}~\Msun/{\rm yr}$, 
where ${\dot M}_{\rm SMBH}$ is the accretion rate onto the SMBH. 
Although cavities around exploding objects may exist in some parameter space, for simplicity, we assume no cavity is present before the explosion in the following analyses.

\begin{figure*}
\begin{center}
\includegraphics[width=180mm]{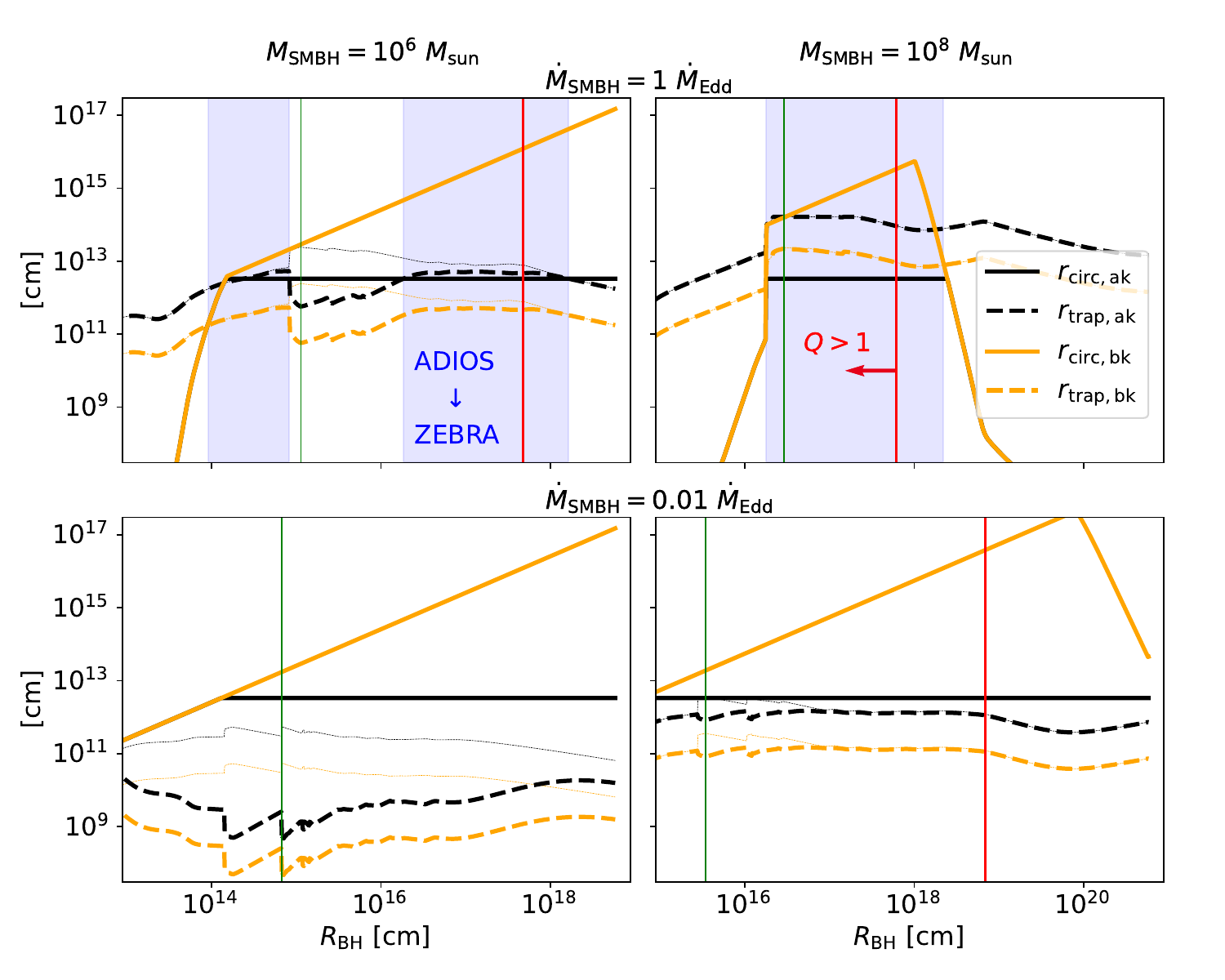}
\caption{
Trapping radius (dashed curves) versus circularization radius  (solid curves) for accretion onto individual BHs in the AGN disk, as a function of distance from the SMBH in the TQM model.
The orange and black lines represent the radii before and after recoil kicks, respectively. 
The left and right panels show results for 
$M_{\rm SMBH}=10^6$ and $10^8~\Msun$, while the upper and lower panels correspond to 
${\dot M}_{\rm SMBH}={\dot M}_{\rm Edd}$ and $0.01~{\dot M}_{\rm Edd}$, respectively. 
Thick lines indicate results considering gap formation around BHs, while thin lines neglect this effect. 
The vertical green and red lines mark the locations 
where 
$H_{\rm AGN}/R_{\rm BH}$ reaches a minimum 
and where the Toomre parameter becomes $Q\geq 1$, respectively. 
The blue shaded regions represent where the ADIOS state transitions to the ZEBRA state as a result of kicks at mergers or interactions. 
}
\label{fig:radii_tqm}
\end{center}
\end{figure*}

\begin{figure*}
\begin{center}
\includegraphics[width=180mm]{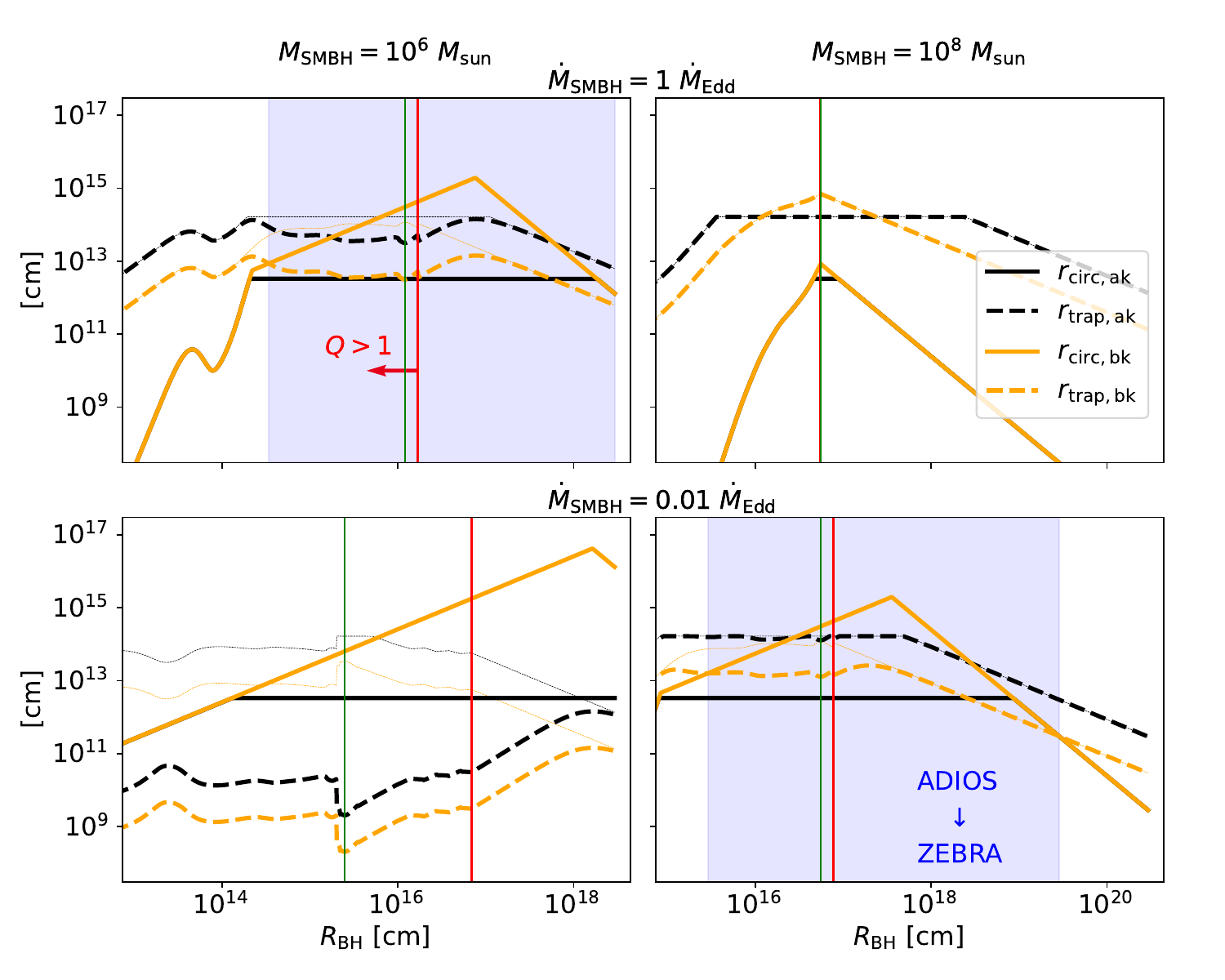}
\caption{
Same as Fig.~\ref{fig:radii_tqm}, 
but for the SG model. 
}
\label{fig:radii_sg}
\end{center}
\end{figure*}

\begin{figure*}
\begin{center}
\includegraphics[width=180mm]{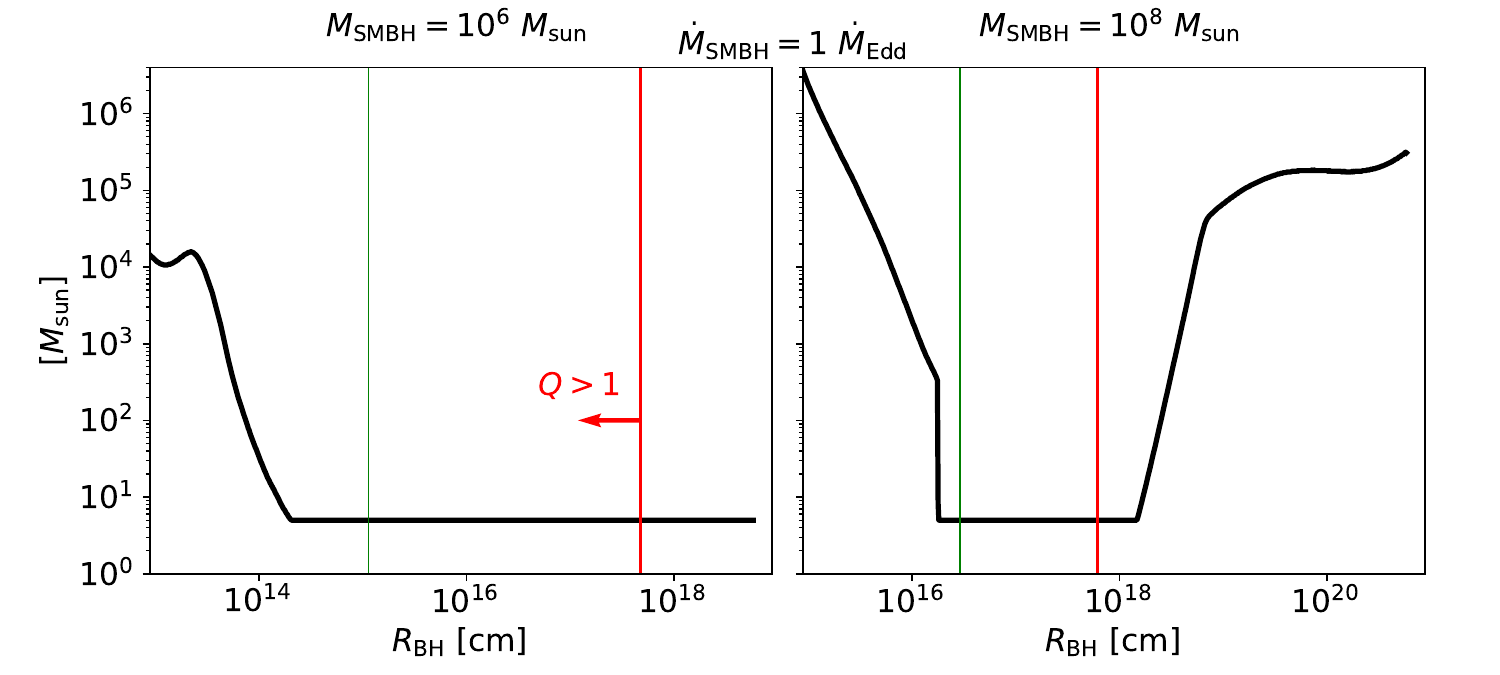}
\caption{
Mass of a BH at which the ADIOS state is realized due to accretion growth prior to kicks, as a function of distance from the SMBH in the TQM model, 
assuming an initial BH mass 
of $m=5~\Msun$. 
For the regions where $m=5~\Msun$, 
the ADIOS state is realized ($r_{\rm circ}>r_{\rm trap}$) without any growth. 
For $m>5~\Msun$, 
the initial state is ZEBRA. 
As the BH rapidly grows, the AGN surface density decreases due to gap formation, reducing $r_{\rm trap}$. Eventually, the ADIOS state is realized ($r_{\rm circ}=r_{\rm trap}$) after growth to the value shown. 
The vertical green and red lines mark the locations 
where 
$H_{\rm AGN}/R_{\rm BH}$ reaches a minimum 
and where the Toomre parameter becomes $Q\geq 1$, respectively. 
}
\label{fig:mass_tqm}
\end{center}
\end{figure*}

\begin{figure*}
\begin{center}
\includegraphics[width=180mm]{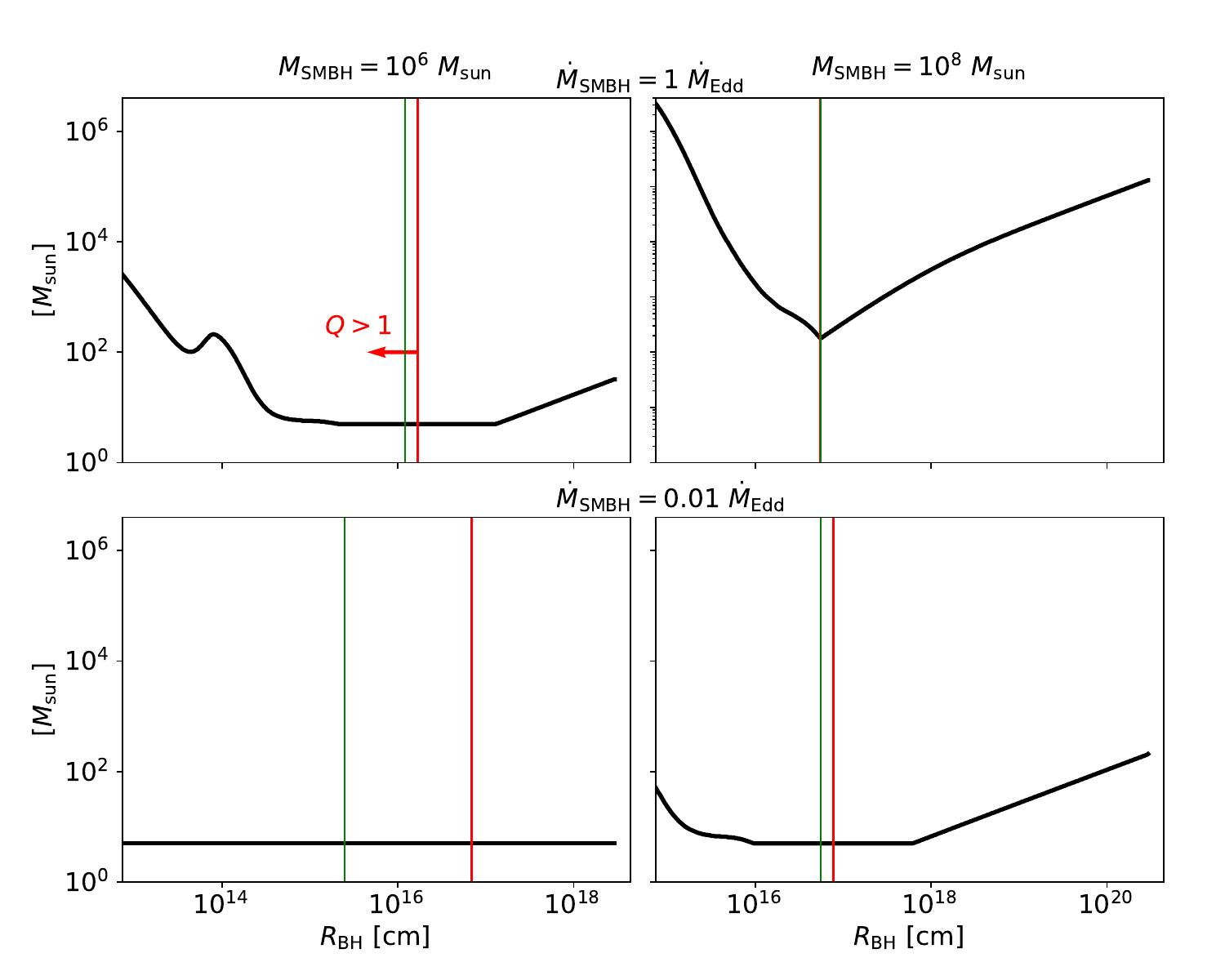}
\caption{
Same as Fig.~\ref{fig:mass_tqm}, 
but for the SG model, and also showing a sub-Eddington AGN (lower panels). 
}
\label{fig:mass_sg}
\end{center}
\end{figure*}

\begin{figure*}
\begin{center}
\includegraphics[width=170mm]{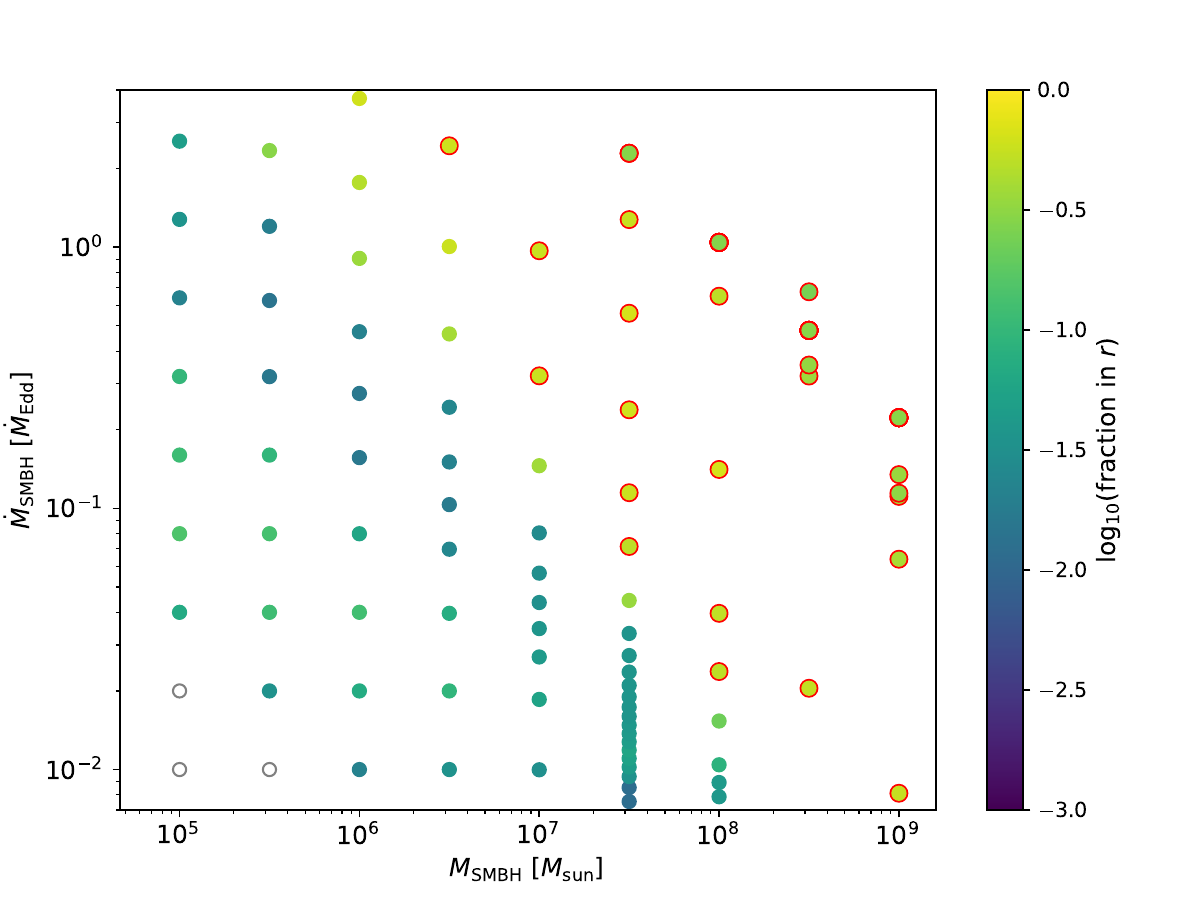}
\caption{
Fraction of the radial distance from the SMBH (logarithmic scale) for cases transitioning from the ADIOS to ZEBRA states 
due to kicks, as a function of the SMBH accretion rate (in units of Eddington rate) and SMBH mass in the TQM model. 
The radial distance ranges from $6\times GM_{\rm SMBH}/c^2$ to $6\times 10^7 \times  GM_{\rm SMBH}/c^2$, with an assumed BH mass $m_{\rm BH}=50~\Msun$. 
Red circles highlight cases where the transition from the ADIOS to ZEBRA state occurs at 
the minimum $H_{\rm AGN}/R_{\rm BH}$ location, 
where mergers are more probable. 
Gray empty circles represent cases where no transition regions appear. 
We evaluate ${\dot M}_{\rm out}$ 
at uniform logarithmic intervals separated by a factor of $2$ for $M_{\rm SMBH}\leq 10^7~{\Msun}$ and $\sqrt{2}$ for $M_{\rm SMBH}> 10^7~{\Msun}$, 
where ${\dot M}_{\rm out}$ is the inflow rate at the outer boundary ($6\times 10^7 \times  GM_{\rm SMBH}/c^2$). 
This results in uneven coverage in ${\dot M}_{\rm SMBH}$, reflecting its nonlinear dependence on ${\dot M}_{\rm out}$. 
In the TQM model, when the AGN luminosity exceeds $\gtrsim 10^{44}~{\rm erg~s^{-1}}$, the transition from the ADIOS to ZEBRA state is associated with kicks at the 
minimum $H_{\rm AGN}/R_{\rm BH}$ region. 
We conclude that high-accretion rate, massive SMBHs are the most promising to produce the EM counterparts as discussed here. 
}
\label{fig:ad_zeb_tqm}
\end{center}
\end{figure*}

\begin{figure*}
\begin{center}
\includegraphics[width=170mm]{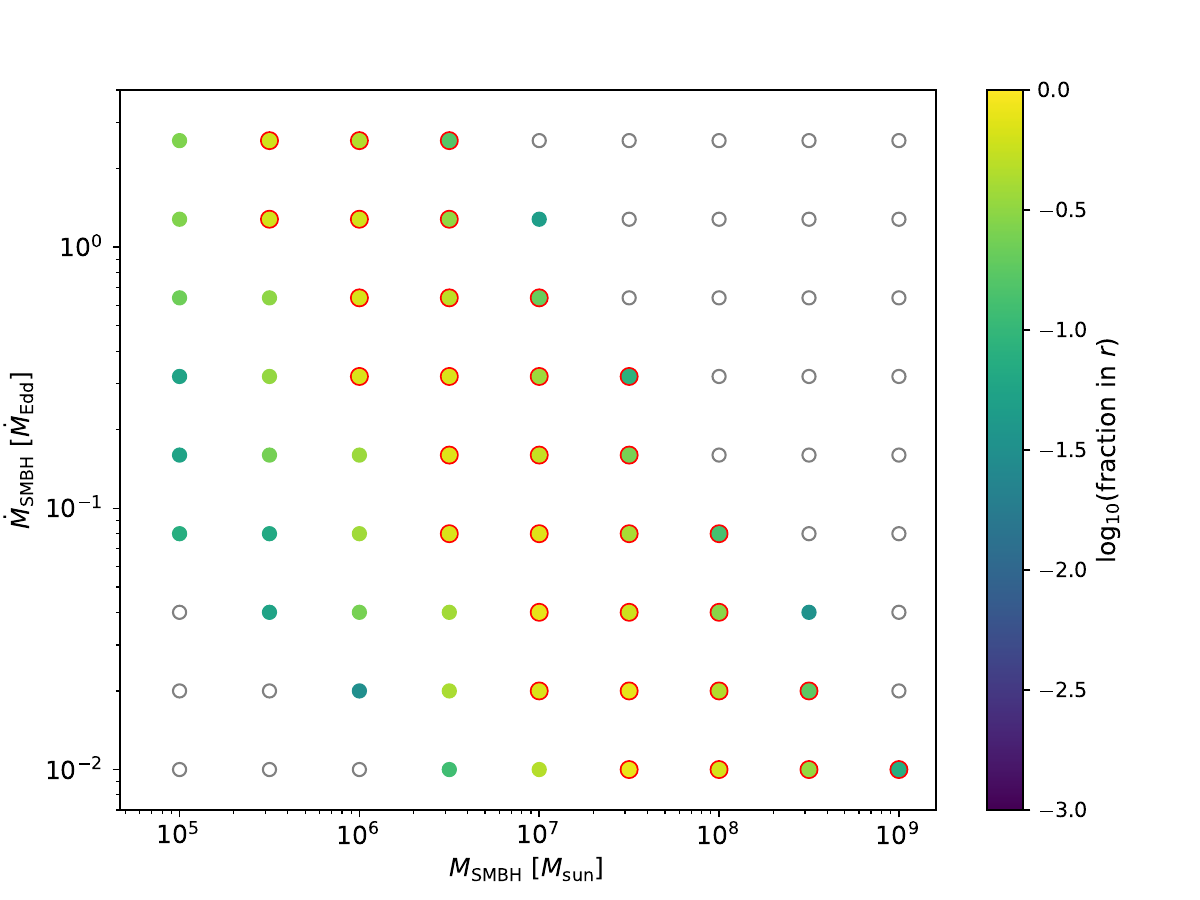}
\caption{
Same as Fig.~\ref{fig:ad_zeb_tqm}, 
shown for the SG model on a uniform logarithmic grid with models separated by a factor of 2. 
In the parameters indicated by gray circles in the lower left and upper right regions, 
BHs with $m_{\rm BH}=50~\Msun$ remain in the ADIOS and ZEBRA states, respectively, irrespective of kicks. 
}
\label{fig:ad_zeb_sg}
\end{center}
\end{figure*}

\begin{figure}
\begin{center}
\includegraphics[width=90mm]{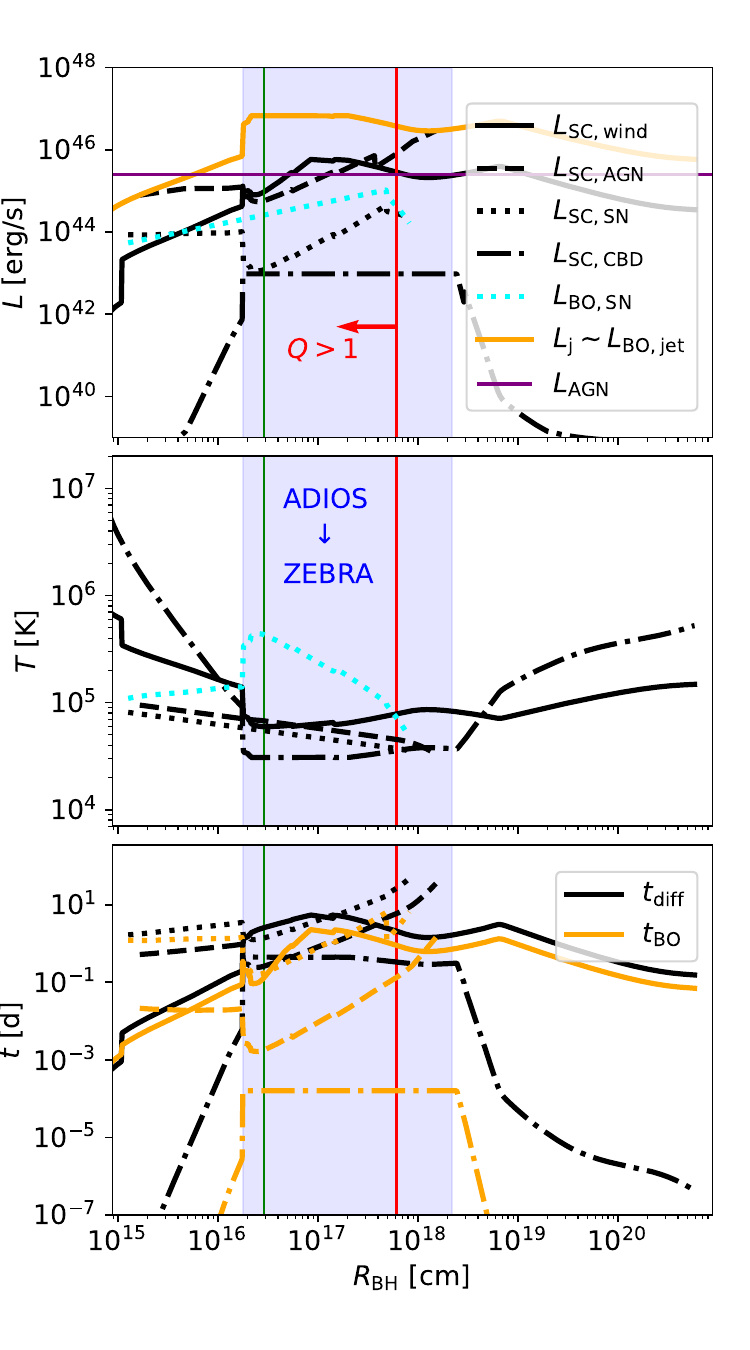}
\caption{
Properties of EM flares in the TQM model 
with $M_{\rm SMBH}=10^8~\Msun$ and ${\dot M}_{\rm SMBH}={\dot M}_{\rm Edd}$. 
Black lines in the upper, middle, and lower rows 
represent the radiation luminosity, the radiation temperature, 
and diffusion timescale for the shock-cooling emission, 
respectively. 
Orange lines in the upper row 
represent the jet kinetic power, 
while those 
in the bottom row represent 
the delay time of shock breakout emission.  
Solid, dashed, and dash-dotted lines 
correspond to emission from shocks in winds, AGN disk gas, and CBDs shocked by jets, respectively, 
while dotted lines represent emission from AGN disk gas shocked by SN explosions. 
Dotted cyan lines in the upper and middle panels 
show the breakout luminosity and temperature for shocks in AGN disk gas caused by SN explosions. 
Vertical green and red lines mark the locations 
where the disk aspect ratio is maximized 
and where the Toomre parameter exceeds 1, respectively. 
The horizontal purple lines represent the AGN luminosity at optical bands, 
assuming a bolometric correction factor of $5$ \citep{Duras2020}. 
The blue shaded regions represent where the ADIOS state transitions to the ZEBRA state as a result of kicks at mergers or interactions. 
The cooling emission from winds and AGN disk gas shocked by jets can be as luminous as the host AGN in the optical-UV bands 
in the middle regions ($\sim 0.01$--$1~{\rm pc}$) with durations of $\sim 0.1$--$10~{\rm d}$. 
}
\label{fig:flare_m}
\end{center}
\end{figure}

\begin{figure}
\begin{center}
\includegraphics[width=90mm]{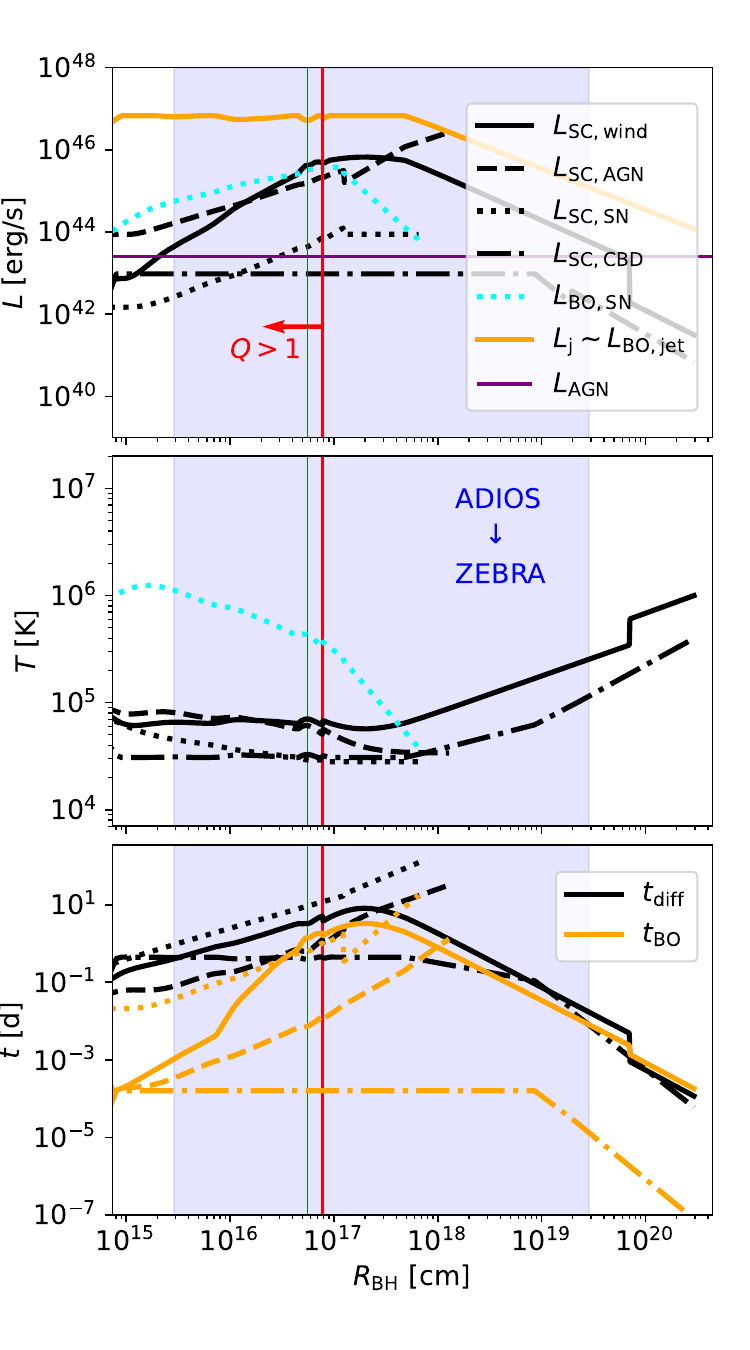}
\caption{
Same as Fig.~\ref{fig:flare_m}, but for the SG model 
with ${\dot M}_{\rm SMBH}=0.01~{\dot M}_{\rm Edd}$ to ensure an ADIOS state for BH accretion 
prior to mergers. 
The cooling emission from winds and AGN disk gas shocked by jets, as well as the emission from AGN disk gas shocked by supernovae, are brighter than the host AGN in the optical-UV bands 
across a wide range of regions ($\sim 10^{-3}$--$10~{\rm pc}$) with durations of $\sim 0.01$--$10~{\rm d}$. 
}
\label{fig:flare_sg}
\end{center}
\end{figure}

\begin{figure*}
\begin{center}
\includegraphics[width=180mm]{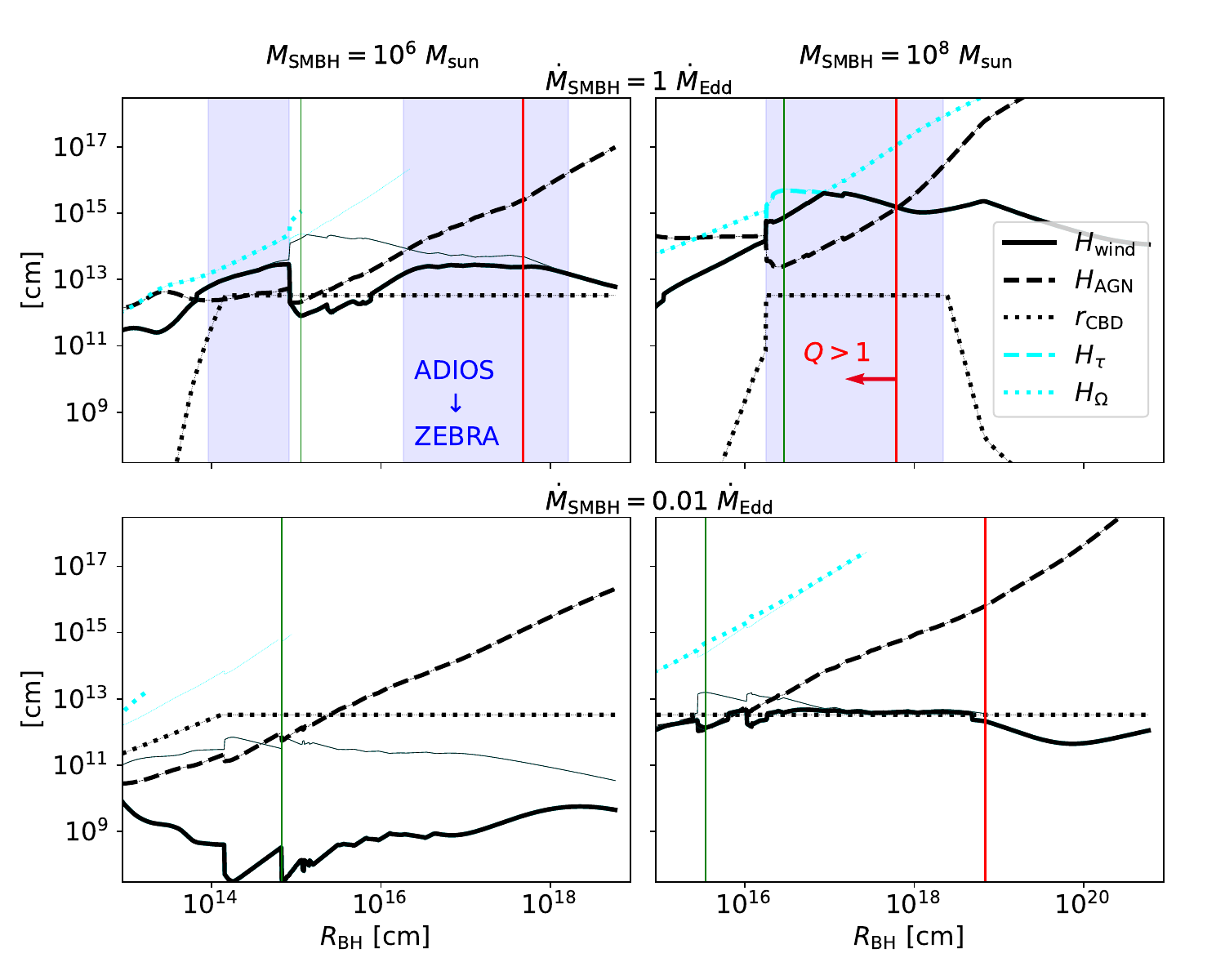}
\caption{
Comparison of various characteristic sizes in the TQM model. 
Solid, dashed, and dotted 
black lines represent the sizes of the wind, AGN disk gas, and CBD, respectively. 
Dashed and dotted cyan lines indicate the wind size limits due to shear from the SMBH ($H_{\Omega}$) and breakout conditions for $\tau>\beta_{\rm ej}^{-1}$
($H_{\tau}$), respectively. 
The lines for $H_{\Omega}$ are plotted only where the local Keplerian velocity about the SMBH exceeds the wind velocity. 
Thick and thin lines denote results 
considering and neglecting gap formation around BHs, respectively. 
Since the shock cooling luminosity is proportional to 
the size of shocked materials (Eq.~\ref{eq:l_sc_rew}), 
these size distributions are useful for understanding the parameters that lead to the production of bright shock-cooling emission. 
The blue shaded regions represent where the ADIOS state transitions to the ZEBRA state as a result of kicks at mergers or interactions. 
The sizes of winds and AGN disk gas are affected by the SMBH mass, the accretion rate onto the SMBH, the radial location, and the AGN disk model. 
}
\label{fig:sizes_tqm}
\end{center}
\end{figure*}

\begin{figure*}
\begin{center}
\includegraphics[width=180mm]{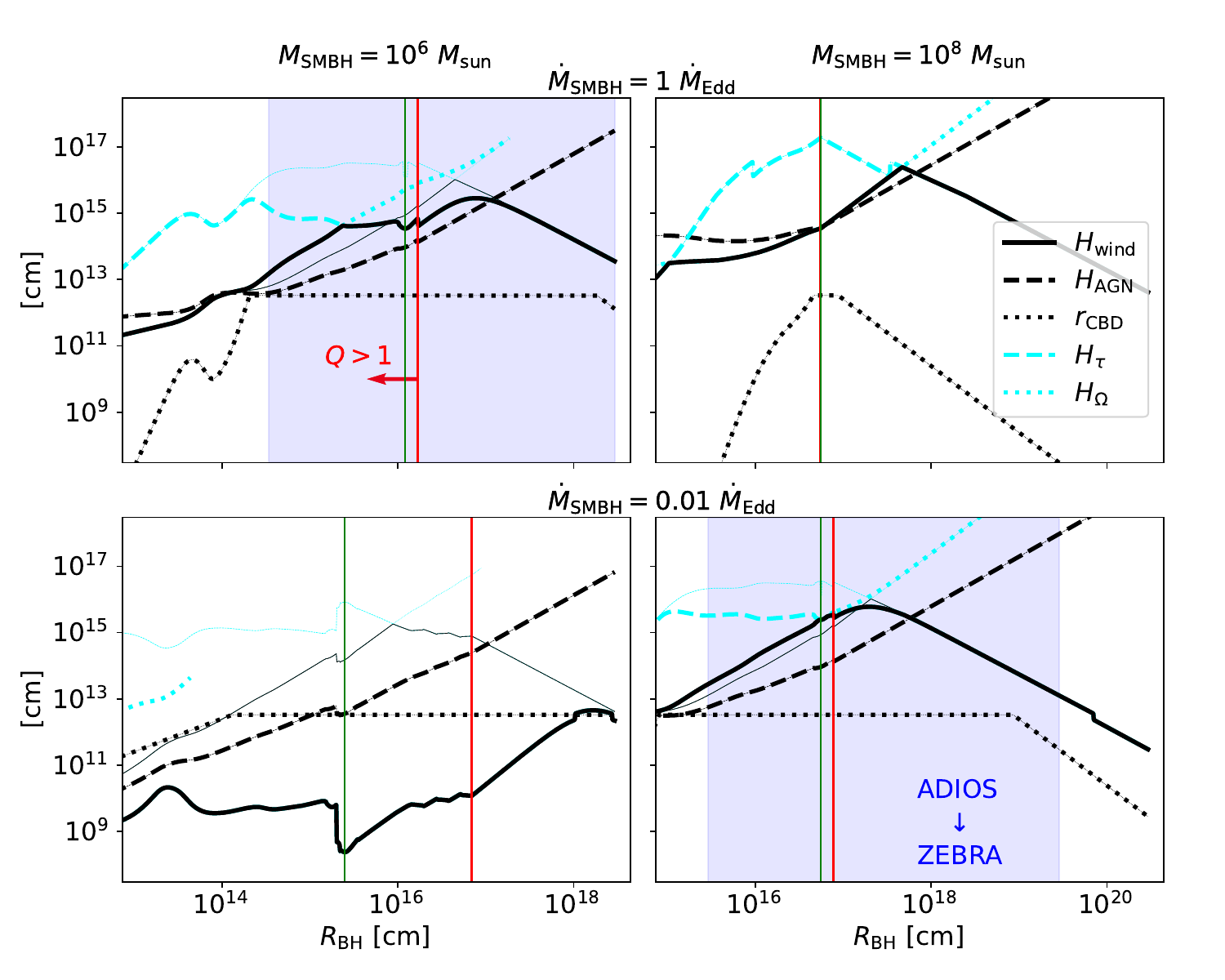}
\caption{
Same as Fig.~\ref{fig:sizes_tqm}, 
but for the SG model. 
}
\label{fig:sizes_sg}
\end{center}
\end{figure*}

\begin{figure*}
\begin{center}
\includegraphics[width=185mm]{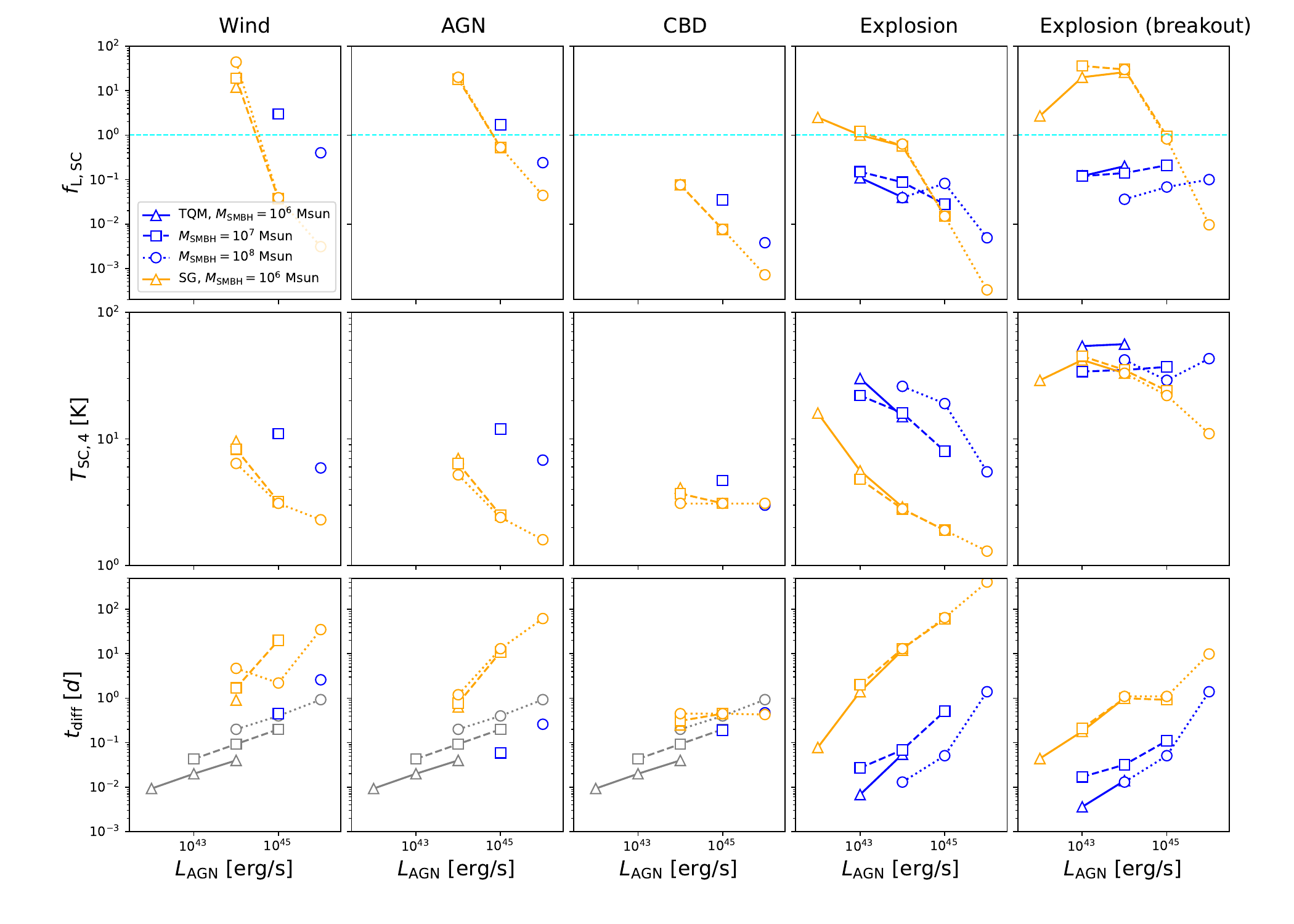}
\caption{
Properties of flares--
luminosity, temperature, and duration--
as functions of AGN luminosity at locations where 
$H_{\rm AGN}/R_{\rm BH}$ reaches its minimum and mergers are more probable. 
The upper, middle, and lower panels show 
the ratio of flare luminosity to AGN luminosity (assuming a bolometric correction of $5$, \citealt{Duras2020}), the radiation temperature in units of $10^4~{\rm K}$, and the diffusion timescale in days, respectively. 
The columns from left to right show results for 
cooling flares from winds, AGN disk gas, and CBDs shocked by jets, as well as 
for cooling and breakout emission from AGN disk gas shocked by SN explosions. 
The lines’ styles correspond to different SMBH masses: solid for $10^6,\Msun$, dashed for $10^7,\Msun$, and dotted for $10^8,\Msun$. 
Blue and orange lines indicate the TQM and SG models, respectively. 
In the lower panels, the delay time between X-ray and optical emission in AGN variability (Eq.~\ref{eq:agn_xo}) is shown with gray lines. 
The dashed cyan line indicates $f_{\rm L,SC}=1$, below which flares are too dim to detect. 
Cooling emission associated with jets from shocked winds and AGN disk gas can be as bright as host AGN emission for $L_{\rm AGN}\sim 10^{44}-10^{45}~\mathrm{erg~s^{-1}}$ in the optical-UV bands. Its properties can be used to distinguish different AGN disk models. 
}
\label{fig:prop_l}
\end{center}
\end{figure*}

\begin{figure*}
\begin{center}
\includegraphics[width=135mm]{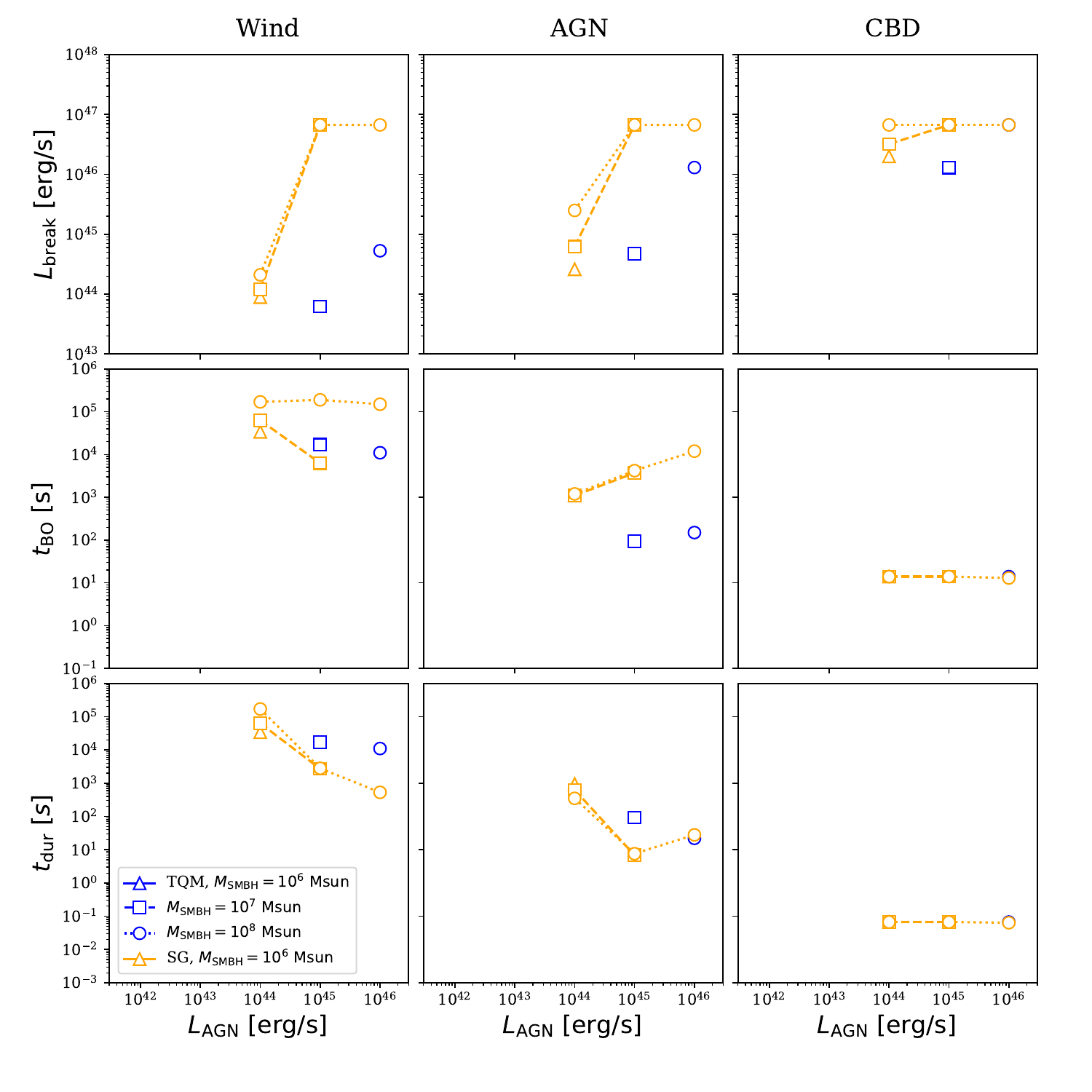}
\caption{
Similar to Fig.~\ref{fig:prop_l}, but for properties of breakout emission associated with jets launched from BHs. 
The upper, middle, and lower panels show the breakout luminosity, duration, and delay time, respectively. 
Since emission from non-jetted AGNs is generally faint in the MeV band compared to these BH-driven flares \citep[e.g.][]{Padovani2017}, the luminosity of BH-driven flares is not compared with AGN luminosity, unlike in Fig.~\ref{fig:prop_l}. 
}
\label{fig:prop_lbo}
\end{center}
\end{figure*}

\section{Results}

\label{sec:results}

In this section, we present our results regarding the accretion modes before and after mergers, 
the typical properties of associated flares, their parameter dependence, and observational strategies for detection.

\subsection{Accretion modes}

\label{sec:results_accretion}

We estimate the accretion states of BHs in AGN disks before and after mergers involving recoil kicks. 
The ordering of the trapping and circularization radii determines which accretion mode (ADIOS or ZEBRA) applies. 
Figs.~\ref{fig:radii_tqm} and \ref{fig:radii_sg} illustrate these radii as functions of the distance from the SMBH, across different SMBH masses and accretion rates in the TQM and SG models. 
The top and bottom panels assume 
SMBH Eddington accretion ratios of ${\dot M}_{\rm SMBH}={\dot M}_{\rm Edd}$ and $0.01~{\dot M}_{\rm Edd}$, respectively\footnote{
In the TQM model, 
inflow rates at outer radii are adjusted to match the accretion rate onto the SMBH. 
}.
The green lines indicate the locations where 
$H_{\rm AGN}/R_{\rm BH}$ reaches minimum, 
often associated with gap formation and increased merger rates \citep{Tagawa19}. 
We consider these locations as potential sites for mergers and 
EM flare production, though mergers can also occur over a wider range, from $\gtrsim 10^{-4}~{\rm pc}$ to several pc \citep{Tagawa2025}.

For $M_{\rm SMBH}=10^6~\Msun$ and ${\dot M}_{\rm SMBH}=0.01~{\dot M}_{\rm Edd}$, 
pre-kick $r_{\rm trap,bk} < r_{\rm circ,bk}$ (orange lines in Figs.~\ref{fig:radii_tqm} and \ref{fig:radii_sg}), 
suggesting wind-suppressed accretion (ADIOS), resulting in faint EM emission.

Conversely, 
in most regions of the SG disk model with 
$M_{\rm SMBH}=10^8~\Msun$ and ${\dot M}_{\rm SMBH}={\dot M}_{\rm Edd}$, 
even prior to the kicks, $r_{\rm trap,bk} > r_{\rm circ,bk}$, 
favoring the ZEBRA mode, rapid growth, and brighter EM emission.

Radiation-pressure-dominated (inner and outer) regions \citep[e.g.][]{Gangardt2024} 
have larger scale heights, reducing the circularization radius 
due to the pressure countering the BH's gravity \citep{Sagynbayeva2024}. 
This favors ZEBRA accretion for $L_{\rm AGN}\gtrsim 10^{43}~{\rm erg~s^{-1}}$ 
(or equivalently ${\dot M}_{\rm SMBH} \gtrsim 0.1~{\dot M}_{\rm Edd} (M_{\rm SMBH}/10^6~\Msun)^{-1}$). 
As the BH grows through ZEBRA accretion, gap formation can reduce the accretion rate and trapping radius, eventually transitioning the mode to ADIOS.

Figs.~\ref{fig:mass_tqm} and \ref{fig:mass_sg} show BH mass thresholds 
for ZEBRA-to-ADIOS transitions caused by surface-density reductions from gap deepening in the TQM and SG models, respectively. 
In these plots, $m_{\rm BH}=5~\Msun$ BHs remain in the ADIOS state, 
which tends to appear in cases of low SMBH masses, low accretion rates onto SMBHs, and in intermediate regions. 
The ADIOS state appears in the entire regions for ${\dot M}_{\rm SMBH}=0.01\times {\dot M}_{\rm Edd}$ in the TQM model, 
which is skipped to be presented in Fig.~\ref{fig:mass_tqm}. 
Transition masses are higher in both inner and outer regions, particularly in luminous AGNs, because the smaller circularization radii favor ZEBRA. Strong mini-jets are anticipated during the ZEBRA phase until the transition to ADIOS \citep{Tagawa2023_highenergy}, which can be observed. Future GW observatories like LISA, TianQin, and Taiji will aid in testing intermediate-mass BH (IMBH) formation in these regions (see also Appendix~\ref{sec:growth}).

After the kicks, 
the condition 
$r_{\rm circ,bk}>r_{\rm b}$ 
can be satisfied for $L_{\rm AGN}\gtrsim 10^{44}$--$10^{45}~{\rm erg~s^{-1}}$ 
(black lines in Figs.~\ref{fig:radii_tqm} and \ref{fig:radii_sg}). 
This condition leads to 
deviations from Keplerian orbits in the outer CBD and generates strong shocks. 
The impact of these shocks on the angular momentum distribution and the resulting transition to the ZEBRA mode depends on hydrodynamical details beyond our simplified model. We assume that, post-kick, the circularization radius becomes $\min(r_{\rm circ,bk},r_{\rm b})$, while the trapping radius increases due to enhanced accretion \citep[e.g.][]{Tagawa2023_highenergy}. 
We assume the transition to ZEBRA occurs 
for $r_{\rm trap,ak} > r_{\rm circ,ak}$, 
with strong emission associated with kicks driven by mergers or binary-single interactions.

Figs.~\ref{fig:ad_zeb_tqm} and \ref{fig:ad_zeb_sg} show the logarithmic fraction of the SMBH radius where accretion-state transitions occur 
for the TQM and SG models, respectively, at the low-aspect-ratio regions. 
The transition from the ADIOS to ZEBRA state occurs 
after kicks 
in middle regions dominated by gas pressure in the AGN disk 
for $L_{\rm AGN}\gtrsim 10^{44}~{\rm erg~s^{-1}}$ in the TQM model and 
$10^{45}~{\rm erg~s^{-1}}\gtrsim L_{\rm AGN}\gtrsim 10^{43}~{\rm erg~s^{-1}}$ for the SG model (red circles). 
If BHs merge in these AGNs, EM counterparts to GW events are expected. 
Differences between the disk models arise from the efficiency of angular-momentum transport, which affects gas density, the BH gas-capture rate, and the trapping radius. 
As AGN luminosity decreases (or increases), the fraction remaining in the ADIOS (or ZEBRA) state increases, 
thereby shaping both the EM signatures and BH growth pathways. 
Future GW detectors, such as LISA, TianQin, and Taiji, will be crucial for constraining the formation of IMBHs in these regions.

\subsection{Properties of flares}

\label{sec:results_prop}

This section examines the luminosity, temperature, and delay time of shock-breakout and shock-cooling emission, 
their dependence on AGN properties, 
and non-thermal emission potentially produced during the breakout phase. 
Figs.~\ref{fig:flare_m} and \ref{fig:flare_sg} 
show these properties for the TQM and SG models, respectively. 
The solid, dashed, and dash-dotted lines represent emission from winds, AGN disk gas, and CBDs shocked by collisions with jets, respectively. 
Dotted lines indicate emission from AGN disk gas shocked by SN explosions. 
The TQM model assumes a high 
accretion rate (${\dot M}_{\rm SMBH}={\dot M}_{\rm Edd}$), 
with post-kick jet production occurring for $M_{\rm SMBH}=10^8~\Msun$, while the SG model assumes a lower accretion rate (${\dot M}_{\rm SMBH}=0.01~{\dot M}_{\rm Edd}$) to ensure a pre-kick ADIOS state for $M_{\rm SMBH}=10^6~\Msun$ 
\footnote{
Emission from shocked AGN disks gas at large radii (Figs.~\ref{fig:flare_m} and \ref{fig:flare_sg}) is not predicted due to the low optical depth ($\tau< c/v_{\rm ej}$), which allows photons to escape immediately, invalidating 
shock breakout assumptions. 
Such explosions are similar to  
typical supernovae (Type~Ia, Ibc, or II) 
with luminosities around $\sim 10^{42}$--$10^{43}~{\rm erg~s^{-1}}$ \citep[e.g.][]{Li2011}
and durations of $\sim 20$--$100~{\rm days}$ \citep[e.g.,][]{Filippenko1997}. 
}.

The properties 
for shock-cooling emission 
depend on the location and on the size and mass of the shocked component
(see Eqs.~\ref{eq:l_sc_rew}, \ref{eq:tem_sc_rew}, and \ref{eq:tdiff_sc_rew} below). 
Figs.~\ref{fig:sizes_tqm} and \ref{fig:sizes_sg} show the sizes of the shocked components as functions of their distance from the SMBH.

The dependence of the flare properties on AGN properties 
is presented in Figs.~\ref{fig:prop_l} and \ref{fig:prop_lbo}.

\subsubsection{Luminosities}

Without the formation of a gap, the jet's kinetic power varies with the BH's radial distance, increasing in the inner regions and decreasing in the outer regions (thin orange lines in the upper panels of Figs.~\ref{fig:flare_m} and \ref{fig:flare_sg}). This variation is influenced by the gas capture rate, 
which is sensitive to the disk's aspect ratio ($\propto R_{\rm BH}^{2}/H_{\rm AGN}^2$ in the absence of gaps). 
The capture rate further diminishes when gap formation occurs. 
Nevertheless, the jet kinetic power reaches extremely high Eddington rates of $\sim 10^5$--$10^7~{\dot M}_{\rm Edd}$ 
for $M_{\rm SMBH}=10^6$--$10^8~\Msun$, 
which would leave observable relics through various processes, as discussed below.

For emission from AGN disk gas, the shock cooling luminosity ($L_{\rm SC}$) increases with increasing $R_{\rm BH}$ (dashed black lines in the upper panels). 
This arises from the relationship:
\begin{align}
\label{eq:l_sc_rew}
L_{\rm SC}=2\pi c v_{\rm ej}^2 R_{\rm BO}/\kappa, 
\end{align}
where $R_{\rm BO}$ is the characteristic size of the shocked gas at jet breakout \citep{Tagawa2023_SC}. 
Notably, $L_{\rm SC}$ is independent of the shocked-gas mass 
because the effects of mass on shock energy, its size during diffusion, and diffusion timescales cancel out. 
Consequently, the luminosity depends only on the ejecta velocity, $v_{\rm ej}$, and the shock size at breakout, $R_{\rm BO}$. 
Since $R_{\rm BO}\propto H_{\rm AGN}$ and $H_{\rm AGN}$ increases with $R_{\rm BH}$, 
(dashed lines in Figs.~\ref{fig:sizes_tqm} and \ref{fig:sizes_sg}), 
and because $v_{\rm ej}$ depends only weakly on $R_{\rm BH}$ 
\footnote{
The ejecta velocity scales roughly as $v_{\rm ej} \propto R_{\rm BH}^{-1/6}$ or $R_{\rm BH}^{-1/10}$, depending on the regime. 
This scaling arises from the relation: 
$v_{\rm ej}\propto (L_j/\rho_{\rm AGN}H_{\rm AGN}^2)^{c} \propto (R_{\rm BH}^{1/2}H_{\rm AGN}^{-1})^c$ with $c=1/3$ and $c=1/5$ for the non-relativistic and relativistic regimes, respectively \citep{Bromberg2011,Tagawa2022_BHFeedback}. 
}, 
the shock cooling luminosity tends to grow with increasing $R_{\rm BH}$.

The shock cooling luminosity from shocked winds ($L_{\rm SC}$)
exhibits a close correlation with the jet luminosity ($L_{\rm j}$) across most of the disk (solid black and orange lines in the upper panels of Figs.~\ref{fig:flare_m} and \ref{fig:flare_sg}). 
This is because the size of wind is limited to $H_{\rm wind}=H_{\tau}\propto {\dot M}_{\rm cap}^{3/2}$, 
particularly in the outer regions
(dashed and dotted cyan lines in Figs.~\ref{fig:sizes_tqm} and \ref{fig:sizes_sg}, Eqs.~\ref{eq:md_wind}, \ref{eq:v_wind}, \ref{eq:h_tau} and \ref{eq:rho_wind}), 
and $L_{\rm SC}\propto R_{\rm BO}\propto H_{\rm wind}$ (Eq.~\ref{eq:l_sc_rew}). 
In the inner regions, 
however, $H_{\Omega}$ is smaller than $H_{\tau}$, especially in the SG model with 
$L_{\rm AGN} \gtrsim 10^{44}~{\rm erg/s}$ 
due to the high ${\dot M}_{\rm cap}$ (Fig.~\ref{fig:sizes_sg}). 
As a result, in these inner regions, $L_{\rm SC}$ can be significantly lower than $L_{\rm j}$.

Regarding the shock-cooling emission from the shocked CBDs, 
$L_{\rm SC}$ remains relatively constant 
over a broad range of $R_{\rm BH}$ 
(dash-dotted black lines in the upper panels of Figs.~\ref{fig:flare_m} and \ref{fig:flare_sg}). 
This arises because their sizes ($R_{\rm BO}$) are primarily limited by the ejection and circularization during the kicks ($r_{\rm b}$, dotted black lines in Figs.~\ref{fig:sizes_tqm} and \ref{fig:sizes_sg}). 
It is important to note that in regions where $L_{\rm SC}$ is not constant, the condition $r_{\rm b} > r_{\rm circ,bk}$ holds, 
indicating that shocks are not caused by kicks. 
In these cases, the jets and associated emission are not caused by kicks.

The luminosity of shock-cooling emission from SN explosions (dotted lines in Figs.~\ref{fig:flare_m} and \ref{fig:flare_sg}), 
depends on the breakout radius $R_{\rm BO}$ similarly to 
the emission from AGN disk gas shocked by jets. 
However, in these cases, 
the luminosity decreases with increasing $R_{\rm BH}$ at larger $R_{\rm BH}$, especially within the SG model. 
This decreasing trend results from the transition from the free-expansion phase to the adiabatic-expansion phase. Beyond this transition, the expansion velocity decreases because the mass enclosed within $R_{\rm BO}$ increases rapidly, causing the shock emission to weaken as the ejecta expand more slowly.

\subsubsection{Temperatures}

\label{section:temperature}

The radiation temperature of shock-cooling emission (black lines in the second rows of Figs.~\ref{fig:flare_m} and \ref{fig:flare_sg}) 
is given by  
\begin{align}
\label{eq:tem_sc_rew}
T_{\rm SC}=(2\pi c^2 R_{\rm BO}/m_{\rm BO}\kappa^2 a)^{1/4},
\end{align}
where $m_{\rm BO}$ is the mass of the shocked gas at breakout, 
and $a$ is the radiation constant. 
For emission originating from the AGN disk gas shocked by jets and SN explosions, 
$T_{\rm SC}$ 
decreases with increasing $R_{\rm BH}$. 
This is because $m_{\rm BO}$ grows faster than $R_{\rm BO}$. 
In gap-forming regions (often appear around the green lines in the left panels), 
$T_{\rm SC}$ is elevated due to a reduction of $m_{\rm BO}$.

In the case of shocks originating from winds, 
the temperature inversely correlates with both the gas-capture rate and $L_{\rm j}$, 
as shown by 
$T_{\rm SC}\propto (R_{\rm BO}/m_{\rm BO})^{1/4}$ (Eq.~\ref{eq:tem_sc_rew}), 
$R_{\rm BO}\propto {\dot M}_{\rm cap}^{3/2}$ 
and $m_{\rm BO}=H_{\tau} {\dot M}_{\rm cap}/v_{\rm wind} \propto {\dot M}_{\rm cap}^3$ 
for the case with $H_{\rm wind}=H_{\tau}$ (Eqs.~\ref{eq:md_wind}, \ref{eq:v_wind}, \ref{eq:h_tau} and \ref{eq:rho_wind}). 
Similarly, for the CBD case, 
it is the case as 
$R_{\rm BO}\sim r_{\rm b}\propto {\dot M}_{\rm cap}^0$ (Eqs.~\ref{eq:rb} and \ref{eq:rkick}) and 
$m_{\rm BO}\propto {\dot M}_{\rm cap}$ (Eq.~\ref{eq:sigma_cbd}).

Similarly, for SN shock breakout,  $T_{\rm BO}$ depends on the local gas density and shock velocity, following 
$T_{\rm BO}\propto \rho_{\rm AGN}^{1/4}v_{\rm ej}^{1/2}$ \citep{Tagawa2023}. 

\subsubsection{Timescales}

The diffusion timescale 
for shock-cooling emission 
(photon diffusion time from inside the shocked material) 
is given by 
\begin{align}
\label{eq:tdiff_sc_rew}
t_{\rm diff}=(\kappa m_{\rm BO}/4\pi c v_{\rm ej})^{1/2}. 
\end{align}
For emission from shocked AGN disk gas, 
$t_{\rm diff}$ generally increases with $R_{\rm BH}$ 
because $m_{\rm BO}$ tends to grow with radius. 
However, in gap-forming regions, $t_{\rm diff}$ decreases due to a smaller 
$m_{\rm BO}$ (black lines in the third row of Figs.~\ref{fig:flare_m} and \ref{fig:flare_sg}). 

For jet-induced shocks, 
$v_{\rm ej}$ varies weakly with $R_{\rm BH}$ because of a constant energy injection. 
In contrast, for SN-driven shocks 
$v_{\rm ej}$ decreases as $m_{\rm BO}$ increases during deceleration phases. 

For shocked winds and CBDs, $t_{\rm diff}$ is correlated with the gas-capture rate, since $m_{\rm BO}$ also depends on it.

The shock-breakout delay time follows a similar trend to $t_{\rm diff}$, as both scale with the size and mass of the shocked region.

\subsubsection{Dependence of shock-cooling emission}

\label{sec:cooling_dependence}

Fig.~\ref{fig:prop_l} 
shows how the properties of 
shock-cooling emission depend on the SMBH mass, 
the accretion rate onto the SMBH, and the AGN disk model. 
It plots $f_{\rm L,SC}$, the radiation temperature in units of $10^4~{\rm K}$  ($T_{\rm SC,4}$), and the diffusion timescale in days ($t_{\rm diff,d}$), at the radii where 
$H_{\rm AGN}/R_{\rm BH}$ reaches its minimum 
(the regions where gaps tend to form and BHs accumulate), 
where $f_{\rm L,SC}$ is the ratio of shock breakout or cooling luminosity to the long-term mean AGN luminosity (assuming a bolometric correction factor in the optical band of $5$; \citealt{Duras2020}).

Shock-cooling emission from winds and AGN disk gas shocked by jets typically has luminosities $\sim 10^{44}$--$10^{45}~{\rm erg~s^{-1}}$, 
and can 
outshine the AGN's intrinsic emission 
when $L_{\rm AGN}\sim 10^{44}$--$10^{45}~{\rm erg~s^{-1}}$. 
In contrast, emission from CBDs shocked by jets and from AGN disk gas shocked by SN explosions generally exhibits lower luminosities, around $\sim 10^{43}$--$10^{44}~{\rm erg~s^{-1}}$. 
These transients typically peak in optical-ultraviolet (UV) bands, with durations and delay times ranging from $\sim 5\times 10^3$ to $10^6~{\rm s}$, depending on the AGN model and luminosity.

Due to the similar dependence of the flare luminosity, temperature, and duration on the SMBH mass and the accretion rate, flare properties can be roughly predicted based on the AGN luminosity, along with the AGN disk model. Consequently, given the AGN luminosity, the flare properties can be used to differentiate between AGN disk models. 
Further discussions on parameter dependence and detection prospects are provided in 
Appendix~\ref{sec:parameter_dependence} and 
$\S~\ref{section:observability}$.  

\subsubsection{Dependence of breakout emission}

\label{sec:prop_breakout}

Fig.~\ref{fig:prop_lbo} summarizes the breakout luminosities, durations, and delays for jet-related breakout emission. 
The jet's kinematic power generally ranges from 
$\sim 10^{46}$ to $3\times 10^{47}~{\rm erg~s^{-1}}$. 
The observed breakout luminosity can be significantly lower than this power if the transparency timescale exceeds the angular and breakout timescales \citep{Tagawa2023}, especially for emission from winds and AGN disk gas in low-$L_{\rm AGN}$ environments. 
Breakout durations and delays span $\sim 0.1$--$10^5~{\rm s}$, primarily determined by the size of the shocked materials. 
Although not considered here, relativistic beaming could further enhance the apparent luminosity and shorten the observed timescales.

Thermal breakout emission from relativistic jet heads 
typically peaks at energies around the MeV scale. 
This is because 
the shocked gas temperature exceeds $\sim 1~{\rm MeV}$ for the shock velocity of $\gtrsim 0.1~c$, 
leading to copious electron-positron pair production. 
These pairs act as an effective thermostat, maintaining the temperature near the MeV band \citep{Nakar2012,Ito2020}. 

Similar to the shock-cooling emission, 
the flare properties can be roughly predicted based on the AGN luminosity, by assuming an AGN disk model. Hence, using the observed AGN luminosity, the flare properties can help distinguish AGN disk models.

SN shock breakout emission from AGN disk gas reaches temperatures of $\sim10^5$ to several times $10^5~{\rm K}$, with durations ranging from $\sim 4\times 10^3$ to $10^5~{\rm s}$. 
In the SG model, 
such breakout emission can outshine the intrinsic AGN emission 
when $L_{\rm AGN}\lesssim 10^{45}~{\rm erg~s^{-1}}$, 
making it potentially detectable in the UV (Fig.~\ref{fig:prop_l}). 
Conversely, in the TQM model, 
the breakout is comparatively faint due to lower AGN densities.

\subsubsection{Non-thermal emission}

\label{sec:nonthermal}

Non-thermal emission may accompany the breakout phase, with durations similar to the thermal emission. 
Its spectrum spans from infrared to gamma-ray energies, 
and it can carry up to $\sim 10\%$ of the jet's kinetic power, depending on plasma conditions \citep{Tagawa2023}. 

If the density profile of the shocked component is very steep, 
radiation-pressure-driven acceleration of the unshocked matter will cease the sharp velocity jump, preventing the formation of shocks that accelerate non-thermal particles via diffusive shock acceleration process \citep[e.g.][]{Kimura2024}. 
On the other hand, more gradual density declines, such as those emerged in winds, 
enable non-thermal particles to be accelerated by shocks, leading to production of non-thermal photons. Observing these non-thermal components can, therefore, provide valuable constraints on the density structure of the shocked material.

\vspace{20pt}

\subsection{Observability of EM flares}

\label{section:observability}

In this section, we assess whether compact-object flares can be detected and identified in observations. 

We focus on two primary types of flares that may originate from compact objects within AGN disks: 
(1) explosive flares, such as SN, and 
(2) flares related to jets associated with BH mergers or binary-single interactions. 
These flares can produce two types of emission phases, as discussed previously: 
(a) shock breakout emission, and 
(b) cooling emission. 
As shown in $\S~\ref{sec:results_prop}$, 
breakout and cooling emission from jet-driven flares are luminous in gamma-ray and optical/UV bands, respectively, while 
those from SN explosions are bright primarily in the UV and optical bands. 
To be detectable and confidently identified, 
these flares must be sufficiently bright and distinguishable from the stochastic variability typically exhibited by AGNs.

\subsubsection{Intrinsic AGN variability}

Numerous studies 
\citep[e.g.,][]{Kelly2009,MacLeod2010,MacLeod2012} model typical AGN variability as a damped random walk (DRW), 
a Gaussian process characterized by a broken-power-law power spectrum or, equivalently, an auto-correlation function that decays exponentially beyond a certain timescale. 
Based on this model \citep{MacLeod2010}, 
for $L_{\rm AGN}=10^{44}$ and $10^{45}~{\rm erg/s}$, respectively, 
the standard deviation of magnitude differences for an infinite time lag is $\sim 0.28$ and $0.13$, 
and the damping timescale is $\sim 190$ and $160~{\rm day}$ for 
$M_{\rm SMBH}=10^8~\Msun$ at the $r$ band (a wavelength of $\sim 600~{\rm nm}$)\footnote{Note that the DRW model overpredicts and possibly underpredicts flare rates on timescales shorter and longer than a few months, respectively \citep{Mushotzky2011,Zu2013,Guo2017}. This may make it easier to discover shorter-duration flares that are not due to intrinsic AGN variability.}. 
With these parameters, 
the probability of observing a magnitude change $\Delta m_r>1~{\rm mag}$ within $30~{\rm day}$ is 
$\lesssim 10^{-4}$ and $\lesssim 10^{-8}$ for $L_{\rm AGN}=10^{44}$ and $10^{45}~{\rm erg/s}$, respectively, 
even if the distribution follows an exponential trend, as suggested by \citet{MacLeod2012} (these flares are even rarer under a Gaussian distribution). 
While DRW variability is thought to originate from processes in the inner-disk, such as X-ray reflection or magnetic heating \citep[][]{Krolik1991,Chauvin2018,Sun2020}, deviations from the DRW model--particularly large, rapid flares--may indicate contributions from other sources, including compact objects. 
Observing flares with a magnitude change 
$\Delta m_r>1~{\rm mag}$ and durations $\lesssim 30~{\rm days}$ in a sample of 
$\lesssim 10^3$ 
AGNs over multiple years could suggest the presence of compact-object flares
\footnote{Note that the DRW model may not reliably predict the rates of very rare flares due to potential non-Gaussian effects that could skew expectations. }. 
We will estimate the occurrence rates of flares with $\Delta m_r\gtrsim 1~{\rm mag}$ or $f_{\rm L,SC}\gtrsim 1$--$2$, caused by jets or explosive events, 
and evaluate their potential to stand out 
from the stochastic variability of AGNs.

Recent observational efforts have focused on identifying peculiar flares that deviate from standard models \citep{Graham20,Graham2023,Ohgami2023,Cabrera2024,Darc2025,He2025,He2025_ZTF23,He2025_GW231123,ZhangHaibin2025,Cabrera2025,Leong2025_spin,Gulati2025,Zhu2026,Bommireddy2026,Vieira2026,Darc2026}, which are invaluable for detecting signatures of compact-object activity.

\subsubsection{Shock breakout emission by jets}

\label{sec:bo_jet}

In this section, we evaluate the detectability of shock breakout emission associated with merging BHs. 
Breakout emission from AGN disk gas shocked by jets 
are expected to be bright in the $\sim {\rm MeV}$ energy bands, 
with the luminosities ranging from $\sim 10^{44}~{\rm erg~s^{-1}}$ to $3\times 10^{47}~{\rm erg~s^{-1}}$ (Fig.~\ref{fig:prop_lbo}). 
This emission has duration spanning from $\sim 0.1~{\rm s}$ to $10^5~{\rm s}$ with delay times between $\sim 10~{\rm s}$ and $10^5~{\rm s}$ in the fiducial (unbeamed) model. 
In the MeV band, emission from non-jetted AGNs is generally faint compared to these BH-driven flares \citep{Padovani2017}, 
so detectability largely depends on the sensitivity of the observing instruments. 
The peak flux is $\sim 9\times 10^{-10}~{\rm erg~s^{-1}cm^{-2}sr^{-1}}(L_{\rm break}/10^{47}~{\rm erg~s^{-1}})(d_{\rm L}/1~\rm {\rm Gpc})^{-2}$, where $d_{\rm L}$ is the luminosity distance to the source. 
The maximum event rate is estimated to be several times higher than the BH-merger rate (see \S~\ref{sec:obs_cooling_jets} below). 
Such events could potentially be detected by the {\it Swift}-Burst Alert Telescope (BAT) instrument \citep{Barthelmy2005}, 
or  
the {\it Fermi} Gamma-ray Burst Monitor (GBM, \citealt{Meegan2009}), 
if they are particularly bright. 
On the other hand, flares with shorter durations of $\sim 0.1$--$100~{\rm s}$, detected at $\gtrsim 100~{\rm keV}$, can be misclassified as gamma-ray bursts, warranting caution. 
Future MeV missions, including the the Compton Spectrometer and Imager \citep{Tomsick2019_COSI}, 
the All-sky Medium Energy Gamma-ray Observatory eXplorer 
\citep{Caputo2022_AMEGO}, 
the Gamma-Ray and AntiMatter Survey \citep{Aramaki2020}, 
eASTROGAM \citep{deAngelis2018}, 
and the Lunar Occultation eXplorer \citep{Miller2019_LOX}, 
could significantly enhance detection capabilities.

\subsubsection{Cooling emission associated with jets}

\label{sec:obs_cooling_jets}

Here we summarize the detectability, durations, luminosities, and multiwavelength features of flares from cooling shocked gas associated with merging BHs in AGN disks, highlighting how they differ from normal AGN variability and other transients.

Our models predict that cooling emission from shocked ambient gas 
is bright in the optical-UV bands 
and can be detectable ($f_{\rm L,SC}\gtrsim 1$) 
for $L_{\rm AGN}\sim 10^{44}$--$10^{45}~{\rm erg~s^{-1}}$, 
originating from shocked winds and AGN disk gas (Fig.~\ref{fig:prop_l}). 
However, detection from shocked CBDs appears more challenging. 
For $L_{\rm AGN}\lesssim 10^{42}~{\rm erg~s^{-1}}$ and $10^{43}~{\rm erg~s^{-1}}$, 
the transition to the ZEBRA state does not occur for $m_{\rm BH}=5~
\Msun$ and $50~\Msun$, rendering cooling emission undetectable. 
The detectable flare durations range from $\sim 1~{\rm day}$ to $\sim~{\rm month}$ in the SG model, 
while in the TQM model, they span from $\sim 0.1~{\rm day}$ to $\sim 1~{\rm day}$. 
These durations scale as $t_{\rm diff}\propto m_{\rm BO}^{1/2}$ (Eq.~\ref{eq:tdiff_sc_rew}), 
and the temperature scales as $T_{\rm SC}\propto m_{\rm BO}^{-1/2}$ (Eq.~\ref{eq:tem_sc_rew}). 
Consequently, both duration and temperature can help distinguish between different AGN disk models. 
Shock-cooling flares with luminosities comparable to 
those of AGNs within $\sim {\rm Gpc}$ can be searched for using optical facilities such as the Zwicky Transient Facility \citep{Bellm_2018}, 
the Vera C. Rubin Observatory \citep{Ivezic2019}, 
the Roman Space Telescope \citep{Spergel_2015}, 
as well as UV satellites 
like the Ultraviolet Transient Astronomy Satellite (ULTRASAT; \citealt{Sagiv2014}), 
the Czech UV satellite mission QUVIK \citep{Werner2024},
and the Ultraviolet Explorer UVEX 
\citep{Kulkarni2021}. 

The BH-merger rate in AGN disks is uncertain; 
the upper limit is roughly constrained by the observed BH merger rate of 
$\sim 10~{\rm Gpc^{-3} yr^{-1}}$ \citep{LIGO2025_O4a_population}. 
Given an AGN density $\sim 10^5~{\rm Gpc^{-3}}$ for $L_{\rm AGN}\gtrsim 10^{44}~{\rm erg~s^{-1}}$ \citep[$z\lesssim 0.5$, ][]{Ueda2014,Duras2020,Shen2020,Ananna2022}, up to $\sim 1$ merger per year may occur per $\sim 10^4$ AGNs. 
\citet{Tagawa19} estimate that 
$\sim 10$ binary-single interactions occur per merger, which could boost the flare rates by a factor of $\sim 10$. 
Monitoring $\sim 10^3$ AGNs over one year at $L_{\rm AGN}\gtrsim 10^{44}~{\rm erg~s^{-1}}$ could potentially detect $\sim 1$ kicked-BH flare. 
This rate is comparable to the rate of random brightening by $\sim 1$ magnitude in roughly 1 in $10^3$ AGNs, as predicted by DRW models over durations of $\sim 30$ days. 
On the other hand, this rate becomes significantly higher compared to the flare rate predicted by DRW models for $\Delta m_r \gtrsim 1.5~{\rm mag}$, durations of $\lesssim 10~{\rm day}$, or $L_{\rm AGN}\gtrsim 10^{45}~{\rm erg/s}$. 
Such flares are great targets for identifying BH-driven events.

Detection requires 
a cadence $\gtrsim 1/t_{\rm diff}$; otherwise, 
longer observation times or larger samples are necessary. 
In the TQM model, 
the low gas density resulting from efficient angular momentum transfer leads to a shorter diffusion timescale (Eq.~\ref{eq:tdiff_sc_rew}). 
This makes high-cadence satellites such as ULTRASAT, with a cadence of $\sim 10~{\rm min}$, particularly effective. 
Coordination with GW alerts can further increase detection prospects.

\citet{Tagawa2023_SC} predicted that a brief X-ray flare may precede an optical flare originating from shock-cooling emission. 
The delay between X-ray and optical signals is comparable to $t_{\rm diff}$, ranging from minutes to months. 
Typical intrinsic AGN variability produces 
an X-ray--optical delay of 
\begin{eqnarray}
\label{eq:agn_xo}
t_{\rm AGN,X-O}\approx 2 &(\lambda/600~{\rm nm})^{4/3} (M_{\rm SMBH}/10^9~\Msun)^{2/3}\nonumber\\
&({\dot M}_{\rm SMBH}/0.1~{\dot M}_{\rm Edd})^{1/3}~{\rm day}
\end{eqnarray}
\citep{Edelson2019,Cackett2021}, where $\lambda$ is the wavelength of the optical emission. 
From the lower panels of Fig.~\ref{fig:prop_l}, 
the delay timescale for AGN variability is similar to that of BH driven flares in the TQM model or emission from CBDs, while it differs from that of emission from winds or AGN disk gas in the SG model. In the former case, information on temperature, magnitude, frequency, and association with GWs is useful for distinguishing their origins. 
If the optical flare results from non-thermal processes \citep{Tagawa2023}, then X-ray and optical emission are expected to brighten simultaneously. 
Therefore, multiwavelength observations are essential, 
as they can help confirm  the origins of the flare and identify distinguishing features.

Compared to 
SNe, tidal disruption events (TDEs), and fast blue optical transients (FBOTs), 
BH-driven flares tend to be more luminous 
and originate at the centers of AGNs, unlike most SNe or FBOTs. 
TDEs typically decline over timescales exceeding $\sim 100~{\rm day}$ following a luminosity decay proportional to $L\propto t^{-\alpha}$ with $\alpha\sim 5/3$, which differs from the faster BH-driven flares\footnote{Note that TDEs exhibit variation in their decay slopes \citep{vanVelzen2021}, 
and there is an unclassified population with properties similar to TDEs but with higher luminosities (Appendix~C of \citealt{Sun2025}).}.
Regarding color evolution, 
BH-driven flares generally exhibit a bluer color that remains roughly 
constant following the Rayleigh–Jeans law, 
while 
$T_{\rm SC}$ evolves above observable bands, similar to TDEs \citep{Ma_2024}. 
Once $T_{\rm SC}$ approaches the observed bands due to adiabatic expansion, 
a color evolves and its evolution becomes modest once hydrogen recombination starts, similar to SNe \citep{Tagawa2023_SC,Faran2019}. 
The temperature keeps decreasing from the rise to the decay phase, resembling SNe \citep{Faran2018} and contrasting with TDEs \citep{vanVelzen2020_uvo}. 
In BH-driven flares, X-ray emission typically precedes optical and is relatively bright, unlike TDEs, SNe, or FBOTs. 
These differences are useful for distinguishing between various transient classes.

\subsubsection{Shock breakout emission from SN explosions}

\label{sec:bo_explosion}

The shock breakout emission from SN shocks propagating through AGN disk gas is predicted to be bright in the UV band. 
It typically lasts from $\sim 4\times 10^3$ to $10^5~{\rm s}$ and can outshine the host AGN 
when $L_{\rm AGN}\lesssim 10^{45}~{\rm erg~s^{-1}}$ in the SG model (Fig.~\ref{fig:prop_l})
\footnote{While the model assumes spherical symmetry, the actual disk geometry 
may influence the observed luminosities \citep{Grishin2021}.}.

If the star formation rate in the AGN disk approximately equals the SMBH accretion rate 
(although it is highly uncertain, e.g., Figs.~5 and 9 of \citealt{Thompson05}), 
and assuming a Salpeter initial mass function with a SN rate per unit mass of $\sim 0.005~\Msun^{-1}$, 
the SN rate per AGN is estimated to be $\sim 10^{-3} (L_{\rm AGN}/10^{45}~{\rm erg~s^{-1}})$. 
This rate exceeds the variability level predicted by the DRW model for luminous AGNs. 
Based on these assumptions, 
monitoring roughly $\gtrsim 10^3$ AGNs over the course of a year with UV satellites ($\S~\ref{sec:obs_cooling_jets}$) could potentially enable the detection of shock breakout emission from SN explosions within AGN disks. 
Since the UV emission from these breakouts in the TQM model tends to be dimmer than typical AGN flares (Fig.~\ref{fig:prop_l}), the detection or absence of such signals could provide constraints on the structure and models of AGN disks.

\subsubsection{Cooling emission from SN explosions}

\label{sec:obs_sc_ex}

Figure \ref{fig:prop_l} indicates that the fraction of cooling emission luminosity, $f_{\rm L, SC}\ll 1$, suggesting that SN shock cooling in AGN disks rarely produces bright flares. 
Exceptions with $f_{\rm L,SC}\gtrsim 1$ occur when $L_{\rm AGN}\lesssim 10^{42}$--$10^{43}~{\rm erg~s^{-1}}$ within the SG model. 
Within a distance of about 300 Mpc—detectable by UV observatories like ULTRASAT—there are roughly $\sim 10^5$ AGNs with $L_{\rm AGN}\sim 10^{42}$--$10^{43}~{\rm erg~s^{-1}}$ \citep{Ueda2014,Duras2020,Shen2020}. Assuming a star formation rate comparable to the SMBH accretion rate (\S\ref{sec:bo_explosion}), continuous monitoring of this large population could yield approximately one observable cooling flare per year. Therefore, detecting cooling emission from SNe in AGN disks via UV wide-field surveys would require monitoring of $\gtrsim 10^5$ AGNs over a year to catch these rare events.

\section{EM counterparts for GW events}

\label{sec:discussion}

In this section, we explore 
possible scenarios for the optical, hard X-ray, and gamma-ray counterparts that have been claimed to accompany BH mergers. 
\citet{Tagawa2023} proposed that thermal and non-thermal shock breakout emission, resulting from collisions between jets and AGN disk gas, can produce gamma-ray and optical flares, respectively, following GW detections. 
Additionally, \citet{Tagawa2023_SC} suggested that shock-cooling emission could also explain the observed optical flares. 
Here, we consider a unified model capable of reproducing 
both gamma-ray, hard X-ray, and optical flares 
using a single consistent set of parameters 
without requiring problematic BH overgrowth. 
Our focus is on shock-breakout emission resulting from collisions between jets and CBDs, as well as shock-cooling emission from collisions between jets and winds or the AGN disks--these represent the most compelling scenarios.

\subsection{High-energy emission}

\subsubsection{Properties of observed flares}

We summarize the observed properties of gamma-ray flares associated with GW events, notably GW150914-GBM and LVT151012-GBM, as discussed in \citet{Connaughton2016} and \citet{Bagoly2016}. 
For GW150914 the {\it Fermi} GBM potentially detected a transient with a luminosity of $\sim 2\times 10^{49}\,{\rm erg~s^{-1}}$, spanning energies from $\sim 10~{\rm keV}$ to several ${\rm MeV}$ energies, with a peak energy at $\sim 2$--$4~{\rm MeV}$. 
This emission was observed 
$\sim0.4\,{\rm{s}}$ after the GW event and lasted $\sim 1~$s. 
The signal-to-noise ratio for this gamma-ray detection was 5.1, with a false alarm probability of 0.0022 (2.9$\sigma$) for association with GW150914 \citep{Connaughton2016}. 
Several studies \citep{Greiner2016, Savchenko2016} raised critiques regarding this detection, which are discussed further in \citet{Connaughton2018}.

A similar transient was reported in association with LVT151012 \citep{Bagoly2016}, with a false alarm probability of 0.04. 
This burst had comparable flux levels and occurred within a short time window of the GW event, with a duration of $\sim 1~{\rm{s}}$ and peak energies between $130~{\rm keV}$ and $3.5~{\rm MeV}$.

Recently, a hard X-ray counterpart to S241125n was reported \citep{2024GCN.38308....1D}. 
The {\it Swift} Burst Alert Telescope (BAT) detected 
a hard X-ray transient in the $15$--$300~{\rm keV}$ range $\sim 11~{\rm s}$ after the GW signal. 
The flare lasted $\sim 0.5~{\rm s}$, with a flux of $1.1_{-0.3}^{+0.2}\times 10^{-7}~{\rm erg~s^{-1}~cm^{-2}}$. 
The spectral index is weakly constrained, ranging from $-0.4$ to $-2$ \citep{2024GCN.38351....1D}. 
The estimated luminosity distance is $\sim 4.2\pm1.6~{\rm Gpc}$ \citep{2024GCN.38313....1L}, and the joint false alarm rate for the spatial and temporal coincidence is 
about once every $\sim 6$ years \citep{2024GCN.38356....1L}. 
From the luminosity distance, the lower limit on the total mass of the merging binary is estimated to be $\sim 120~\Msun$ \citep{Zhang2025}.

Follow-up X-ray observations of the S241125n-BAT flare using
the Follow-up X-ray Telescope 
on board the Einstein Probe
identified a possible candidate in the $0.5$--$10~{\rm keV}$ range, with a flux of $\sim 1.2\times 10^{-13}~{\rm erg~s^{-1}~cm^{-2}}$, detected $\sim$26 hr after the GW event within the Swift/BAT localization region (within a 5 arcminute circle). 
However, 
multiple X-ray sources are present within this larger localization area (within a 10 arcminute circle, \citealt{2024GCN.38345....1W}).

\subsubsection{Intrinsic parameters}

\label{sec:results_gamma}

We now aim to constrain the physical parameters necessary to reproduce the observed properties of these gamma-ray and X-ray flares, 
assuming they originate from shocks produced by collisions between a CBD and a jet reoriented at merger. 
This scenario is considered to account for the observed short delays and durations, 
which are challenging to account for through emission from shocked AGN gas and winds.

The Lorentz factor of the jet head for a collimated jet is approximated as \citep{Bromberg2011}
\begin{align}
\label{eq:gamma_cbd1}
\gamma_{\rm h} \simeq {\rm max}\left\{
1,\left( \frac{L_{\rm j}}{32\rho_{\rm CBD} r_{\rm CBD} \theta_0^4 c^3}  \right)^{1/10} \right\}. 
\end{align}
Using relations between the jet kinetic power and the accretion rate (Eq.~\ref{eq:lj_macc}), 
between the accretion rate and the CBD density (Eq.~\ref{eq:sigma_cbd}), 
and formula for the CBD size (Eq.~\ref{eq:rb}), 
this simplifies to
\begin{eqnarray}
\label{eq:gamma_h_sim} 
\gamma_{\rm h}&=&\left(\frac{3 \pi \eta_j \alpha v_{\rm kick}}{16 \theta_0^4 c f_{\rm b}^{1/2}}\right)^{1/10}\nonumber\\
&\approx&1.7
\left(\frac{\eta_j \alpha f_{\rm b}^{-1/2}}{0.2}\right)^{1/10} 
\left(\frac{\theta_0}{0.05}\right)^{-2/5}
\left(\frac{v_{\rm kick}}{500~{\rm km~s^{-1}}}\right)^{1/10}. 
\end{eqnarray}
The Lorentz factor of the forward shock is $\gamma_{\rm FS}\simeq \sqrt{2}\gamma_{\rm h}$ and the final Lorentz factor of the shocked gas is $\gamma_{\rm h,f}\simeq \gamma_{\rm h}^{1+\sqrt{3}}\simeq 4.3$ for $\gamma_{\rm h}\sim 1.7$. 
This $\gamma_{\rm h,f}$ 
is roughly consistent with 
the observed peak energy of GW150914-GBM ($\sim 2$--$6~{\rm MeV}$), 
considering Doppler boosting and 
pair-annihilation temperature constraints \citep{Katz2010,Ito2018,Ito2020}. 

The delay between the EM and GW signals (Eq.~8 of \citealt{Tagawa2023}) is 
\begin{align}
\label{eq:t_delay} 
t_{\rm delay}&=\frac{r_{\rm b}}{4\gamma_{\rm FS}^2 c} \nonumber\\
&\approx 0.9
\left(\frac{\eta_j \alpha}{0.08}\right)^{-1/5} 
\left(\frac{\theta_0}{0.05}\right)^{4/5}\left(\frac{v_{\rm kick}}{500~{\rm km~s^{-1}}}\right)^{-11/5}\nonumber\\
&\times
\left(\frac{m_{\rm BH}}{60~{\Msun}}\right) 
\left(\frac{f_{\rm b}}{0.2}\right)^{11/10}~{\rm sec}. 
\end{align}
Similarly, the duration of the flare (Eq.~11 in \citealt{Tagawa2023}) is 
\begin{align}
\label{eq:t_dur} 
t_{\rm dur}&=\frac{r_{\rm b}}{2\gamma_{\rm h,f}^2 c} \nonumber\\
&\approx 0.6(r_{\rm b}/6\times 10^{11}~{\rm cm})(\gamma_{\rm h,f}/4.3)^{-2}~{\rm sec}. 
\end{align}

Assuming the shock breakout luminosity $L_{\rm break}=f_{\rm beam}L_{\rm j}$, 
where the beaming factor is estimated as $f_{\rm beam}\sim 2 \gamma_{\rm h,f}^2$, and incorporating relations for the jet power and accretion rate (Eq.~\ref{eq:lj_macc}), 
the accretion rate and trapping radius (Eq.~\ref{eq:r_trap}), 
the accretion rate and the kick/trapping radii (Eq.~\ref{eq:macc_af}), 
and the kick radius with the kick velocity (Eq.~\ref{eq:rkick}), 
the expression for $L_{\rm break}$ becomes
\begin{align}
\label{eq:l_gamma_val} 
L_{\rm break}&=
2\gamma_{\rm h,f}^2 \eta_{\rm j} c^2 {\dot M}_{\rm acc,ak}
\nonumber\\
&=
2\gamma_{\rm h,f}^2 \eta_{\rm j} L_{\rm Edd} f_{\rm inc}
(c/v_{\rm kick})^{2p} ({\dot m}_{\rm cap,bk})^{1-p} 
\nonumber\\
&\approx 4\times 10^{48}
\left(\frac{f_{\rm inc}}{60}\right)
\left(\frac{\eta_{\rm j}}{0.8}\right) 
\left(\frac{m_{\rm BH}}{60~{\Msun}}\right) \nonumber\\
&\times\left(\frac{v_{\rm kick}}{500~{\rm km~s^{-1}}}\right)^{-2}
\left(\frac{\gamma_{\rm h,f}}{4.3}\right)^{2} 
~{\rm erg~s^{-1}} 
\end{align}
for $p=1$ and $r_{\rm kick}\leq r_{\rm trap,bk}$. 
With the parameters adopted in Eq.~\eqref{eq:l_gamma_val}, 
$L_{\rm break}$ increases by a factor of $2.4$ for $p=0.5$. 
Its dependence is sensitive to the ratio $r_{\rm kick}/r_{\rm trap,bk}$ as $L_{\rm break}\propto (r_{\rm kick}/r_{\rm trap,bk})^p$. 
\citet{Connaughton2018} argued that observed 
gamma-ray luminosities might be overestimated; 
thus, the calculated $L_{\rm break}$ (Eq.~\ref{eq:l_gamma_val}) can be compatible with observations when uncertainties are taken into account.

The parameters used in these estimates are: 
recoil velocity, $v_{\rm kick}\sim 500~{\rm km~s^{-1}}$ 
(due to mergers, \citealt{Buonanno08}, or binary-single interactions, \citealt{Tagawa19}), 
accretion enhancement factor $f_{\rm inc}=60$ (Appendix~B of \citealt{Tagawa2023}), 
and opening angle of an injected jet, 
$\theta_0 \sim 0.05~{\rm rad}$ \citep{Hada2018,Berger2014}, 
which are consistent with theoretical expectations and observational constraints.

For the event S241125n-BAT, 
adopting $\theta_0=0.02~{\rm rad}$, $v_{\rm kick}=300~{\rm km~s^{-1}}$, and $m_{\rm BH}=150~\Msun$, 
and using values for the other parameters consistent with GW150914-GBM,  
the estimated timescales and luminosity are: $t_{\rm dur} \simeq 0.6~{\rm s}$, $t_{\rm delay}\simeq 3~{\rm s}$, and $L_{\rm break}\simeq 2\times 10^{50}~{\rm erg~s^{-1}}$. 
These values are roughly consistent with the observational data. 
The observed delay of $\sim$11 s can be explained by an additional jet-launch delay ($\lesssim 10^4~r_{\rm g}/c \sim 7~{\rm s}$, e.g., \citealt{Curd2023}). 
Furthermore, a high Lorentz factor of the shocked gas $\gamma_{\rm h,f}\sim 10$ could account for the short duration $t_{\rm dur}\ll t_{\rm delay}$.

The apparent peak energy depends on Doppler boosting and 
the photon-to-baryon ratio ${\tilde n}$. 
Transitions between photon-starved and photon-rich breakout regimes occur near ${\tilde n}\sim 10^3$ \citep{Ito2018}. 
For CBDs with high accretion rates (${\dot m}_{\rm Edd}\sim 10^6$) and $m_{\rm BH}=150~\Msun$, the estimated ${\tilde n}$ is about $\sim 2\times 10^3$ (increasing with accretion rates). This places the system near the transition between regimes, 
implying that the peak photon energies could range from $\sim {\rm MeV}$ to $\sim 10^5~{\rm eV}$ for ${\tilde n}=10^3$ to $10^4$. 
Variations in the spectral peak among different events could offer valuable constraints on ${\tilde n}$ and the Lorentz factor ($\gamma_{\rm h,f}$). 
Consequently, observations of the spectral peak can thus help elucidate the radiation processes involved in breakout emission.

The faint X-ray source detected roughly 26 hours after the GW event, reported by the Einstein Probe \citep{Zhang2025}, may represent non-thermal emission arising from ejecta interacting with the surrounding medium. Further modeling is required to confirm this scenario, but it presents an intriguing avenue for future investigations.

\subsubsection{Delay time for gamma-ray flares}

\label{section:delay_time}

Prompt 
gamma-ray flares with delays $\lesssim {\rm s}$
require 
the establishment of the ZEBRA accretion state even before the merger. 
The ZEBRA mode enhances accretion at least on the dynamical timescale at $r_{\rm b}$, estimated as: 
$t_{\rm enh}\sim 3\times 10^4(v_{\rm kick}/300~{\rm km~s^{-1}})^{-3} (f_{\rm b}/0.2)^{3/2} (m_{\rm BH}/60~\Msun)~{\rm s}$.
To ensure the accretion rate is sufficiently enhanced at the time of merger, 
the merger needs to occur 
(i) before the CBD is depleted within $r_{\rm b}$, and 
(ii) after the transition to the ZEBRA state. 

A promising channel to realize the ZEBRA state at merger 
is GW capture (GWC) via binary-single interactions \citep[e.g.][]{Samsing14}. 
In this process, 
chaotic interactions and shocks can reduce $r_{\rm circ}$, driving the inner disk into the ZEBRA mode prior to merger, leading to jet formation. 
At merger, the jet's direction is reoriented, causing it to collide with the CBD and produce a gamma-ray flare beamed toward the observer. 
In this scenario, 
the merger occurs rapidly before the CBD is significantly depleted, and 
the reoriented jet then breaks out in a different direction, colliding with surrounding material and resulting in emission shortly after the merger.

Several relevant timescales are involved: 
Binaries can merge through GWC at roughly $\sim 100$ times the dynamical timescale of the binary (prior to interaction, $t_{\rm dyn}$) after the start of binary-single interactions ($t_{\rm BS-mer}$), 
and 
$\sim 10$ times $t_{\rm dyn}$ after the last chaotic three-body interaction ($t_{\rm int-mer}$) \citep{Samsing14}. 
For a binary with semi-major axis $a_{\rm bin}\sim 10^{11}~{\rm cm}$ (typical for AGN-disk binaries, \citealt{Tagawa19}), 
the merger time after the final interaction is $t_{\rm int-mer}\sim 10~t_{\rm dyn}\sim 2\times 10^4(a_{\rm bin}/10^{11}~{\rm cm})^{3/2}(m_{\rm BH}/60~\Msun)^{-1/2}~{\rm s}$. 

To enable the enhancement of accretion prior to merger, we require 
(i) $t_{\rm BS-mer}>t_{\rm enh}$. 
Moreover, to prevent significant gas depletion before merger, 
(ii) the timescale $t_{\rm int-mer}$ needs to be shorter than the viscous timescale at $r_{\rm b}$, where  
\begin{eqnarray}
\label{eq:t_vis}
t_{\rm vis}&\sim& \left(\frac{r_{\rm b}^3}{Gm_{\rm BH}}\right)^{1/2}\left(\frac{1}{\alpha h_{\rm CBD}^2}\right)\nonumber\\
&\sim& 3\times 10^5
\left(\frac{f_{\rm b}}{0.2}\right)^{3/2}
\left(\frac{v_{\rm kick}}{300~{\rm km~s^{-1}}}\right)^{-3}\nonumber\\
&&\left(\frac{m_{\rm BH}}{60~{\Msun}}\right)
\left(\frac{\alpha h_{\rm CBD}^2}{0.1}\right)^{-1}~{\rm sec}.
\end{eqnarray}
Using typical parameters, these conditions (i) and (ii) are satisfied, allowing the reoriented jet to rapidly collide with 
a quasi-spherical CBD, 
producing bright emission shortly after the merger.

Jets generated before merger will 
interact with pre-existing extended winds, 
which are replenished on timescales roughly: $t_{\rm wind} \sim H_{\rm wind}/v_{\rm wind}\gtrsim 10^6~{\rm s}$ (for $M_{\rm SMBH}=10^6$--$10^8~\Msun$ in low-aspect-ratio regions). 
This is much longer than the merger timescale via the GWC mechanism,
so the emission from the shocked CBD remains largely unobscured and unscattered. 
The quasi-spherical CBD 
within the cavity (opening angle $\theta_0$) refills on roughly a dynamical timescale at $r_{\rm b}$ ($\sim 5\times 10^3(r_{\rm b}/5\times 10^{11}~{\rm cm})^{3/2}(m_{\rm BH}/60~\Msun)^{-1/2}~{\rm s}$), 
which means the re-oriented jet has a high probability of colliding with this 
CBD, producing observable emission.

This scenario also predicts that gamma-ray flares could occur 
before the merger, with a typical time difference, $t_{\rm int-mer}\sim 2\times 10^4~{\rm s}$. 
Detecting such pre-merger flares would serve as an important test of this model. 
For first-generation BHs (e.g., GW150914 progenitors), the spins are generally low ($a\lesssim 0.3$, \citealt{Abbott16a,Abbott2023_O3_Properties}), 
suggesting that jets before merger may be weak ($\eta_j \propto \sim a^2$, \citealt{Tchekhovskoy2011}), 
making pre-merger flares more challenging to detect. 
Conversely, if the merging BHs are remnants of previous merger events (e.g., GW190521 \citealt{LIGO20_GW190521_astro}), stronger jets and luminous emission become plausible. 

Finally, optical flares arising from shock-cooling emission generated by AGN disk gas or winds shocked before mergers can become observable following the arrival of GW signals. This is because the diffusion timescales typically exceed the merger timescale via the GWC mechanism ($t_{\rm BS-mer}$). 
A joint detection of gamma-ray and optical flares would serve as a robust test of this scenario.

\subsubsection{Association probability}

\label{sec:prob_gamma}

We estimate the probability that high-energy (gamma-ray or X-ray) flares are observed to be associated with GW events. 
Note that this probability differs from the detection rates estimated without considering GW observations in $\S~\ref{section:observability}$.

The rapid appearance of gamma-ray flares following GW detection 
suggests that mergers are likely facilitated by the GWC mechanism (\S~\ref{section:delay_time}). 
The fraction of mergers occurring via GWC after binary-single interactions is uncertain; estimates range from $P_{\rm GWC}\sim 5\%$ to $90\%$ for mergers in AGN disks \citep{Tagawa20_ecc,Samsing20,Rowan2025}. 
We adopt a fiducial value of $P_{\rm GWC}\sim 0.1$,  representative of isotropic binary-single interactions without gas effects. 

The probability of detecting beamed emission is reduced by a factor of 
$P_{\rm beam}\sim {\rm max}(\gamma_{\rm h,f}^{-2},\theta_0^2)$. 
Assuming a Lorentz factor of a few, 
we adopt $P_{\rm beam}\sim 0.1$.

We further assume that the gamma-ray flux from shocks in the CBD exceeds the AGN background with probability $P_{\rm bright}\sim 1$ whenever the emission is beamed toward us.

Referring to the TDE simulations \citep{Steinberg2024,Price2024}, 
circularized CBDs form a quasi-spherical shape before cooling timescale (which is longer than the accretion timescale for the ZEBRA flow, given $r_{\rm circ}<r_{\rm trap}$), 
surrounding the BH in most directions. 
We therefore 
adopt a collision probability $P_{\rm coll}\sim 1$.

The active duty cycle of BHs in AGN disks 
may be reduced by cavity formation caused by winds. 
\citet{Tagawa2022_BHFeedback} estimate that this cycle ranges between $P_{\rm active}\sim 0.1$--$1$. 
We adopt $P_{\rm active}\sim 1$, which is appropriate for gap-forming regions.

Combining these factors, 
the optimistic probability that gamma-ray flares from BH mergers in AGN disks are observable 
is 
\begin{align}
P_{\rm association}&\sim P_{\rm GWC} P_{\rm beam} P_{\rm col} P_{\rm bright} P_{\rm active} P_{\rm typeI}\nonumber\\
&\sim 0.01
\left(\frac{P_{\rm GWC}}{0.1}\right) 
\left(\frac{P_{\rm beam}}{0.1}\right) 
\left(\frac{P_{\rm col}}{1}\right) \nonumber\\
&\times\left(\frac{P_{\rm bright}}{1}\right) 
\left(\frac{P_{\rm active}}{1}\right). 
\end{align}

Considering partial sky coverage and downtime, the detection probability with
the {\it Fermi} Gamma-ray Burst Monitor \citep{Meegan2009} 
further reduces this estimate by a factor of 
$\sim 3$. 

In conclusion, gamma-ray flares associated with GW events are expected to be rare—on the order of $\sim 10^{-2}$—even with multiple telescopes. 
Although the uncertainty is large, the current model can still account for the suggested associations of the gamma-ray events, including GW150914-GBM, LVT151012-GBM, and S241125n-BAT flares. 
A larger GW sample will be crucial for testing these scenarios.

\subsection{Optical emission}

\subsubsection{Properties of observed flares}

\label{sec:optical_observation}
Seven optical flares reported 
by \citet{Graham20} and \citet{Graham2023} 
began to exceed the persistent AGN flux $20$--$200~{\rm days}$ (rest frame) after the GW merger. 
Their peak luminosities in the $g$ and $r$ bands are $\sim 10^{44}$--$10^{45}\,{\rm erg~s^{-1}}$, with durations of $\sim 20$--$100\,{\rm days}$ (rest frame). 
The host SMBH masses are estimated to lie between $M_{\rm SMBH}\sim 10^8$ and $10^9\,\Msun$, 
and their Eddington ratios range from $\sim 0.02$ to $\sim 0.2$ \citep{Graham20,Graham2023}.

\subsubsection{Intrinsic parameters}

\label{sec:results_optical}

We assume that 
the shock-cooling emission properties 
are given by $L_{\rm SC}=10^{46}~{\rm erg~s^{-1}}$, $T_{\rm SC}=4\times 10^4~{\rm K}$, and $t_{\rm diff}=30~{\rm day}$, 
which roughly reproduce the luminosity in the optical bands (wavelengths of $\sim 500$--$700~{\rm nm}$) of $\sim 1$--$3\times 10^{45}~{\rm erg/s}$ ($\S~\ref{sec:optical_observation}$). 
Using these values, we derive the intrinsic parameters as 
$R_{\rm BO}=6\times 10^{15}~{\rm cm}$, $m_{\rm BO}=5~{\Msun}$, and $v_{\rm ej}=0.07~c$, 
using the relations between the observable properties and the shocked gas parameters  (Eqs.~\ref{eq:l_sc_rew}, \ref{eq:tem_sc_rew}, and \ref{eq:tdiff_sc_rew}). 
If $v_{\rm ej}\gg 0.1c$, the shocked gas cannot be thermalized and cannot produce bright optical emission (see \S~3.4 of \citealt{Tagawa2023_SC}), which provides a testable constraint on the model.

With these assumptions, we can estimate the accretion rate prior to the transition in the accretion state 
for emission originating from shocked winds. 
Using $\rho_{\rm wind}\sim M_{\rm BO}/4\pi R_{\rm BO}^3$ 
and combining this density with the equations governing wind production rate, wind velocity, and wind density (Eqs.~\ref{eq:md_wind}, \ref{eq:v_wind}, and \ref{eq:rho_wind}), 
the gas capture rate can be estimated as 
\begin{eqnarray}
\label{eq:m_wsh}
{\dot m_{\rm cap,bk}}&=&
\left(\frac{\Omega_{\rm w} M_{\rm BO}  \theta_0^{-4/3} c}{4\pi R_{\rm BO} {\dot M}_{\rm Edd} \eta_{\rm rad} }\right)^{2/3}, 
\end{eqnarray}
where we assume $H_{\tau}\simeq R_{\rm BO}\theta_{0}^{-2/3}$, considering the geometry of the shocked gas for collimated jets \citep{Bromberg2011}. 
Using the above parameters, with $m_{\rm BH}=100~\Msun$ and $\theta_0=0.05$, 
the estimated capture rate is ${\dot m_{\rm cap}}\sim 3\times 10^{7}$.

Similarly, the jet kinetic power for emission originating from shocked AGN gas can be estimated. 
Using $\rho_{\rm AGN}\sim M_{\rm BO}/4\pi R_{\rm BO}^3$, 
and the scale height of the AGN disk estimated via 
$R_{\rm BO}^3 \sim H_{\rm AGN}^3 \theta_0^2 f_{\rm corr}^3$, 
based on Eq.~(1) of \citet{Tagawa2022_BHFeedback}, 
the Eddington ratio for the capture rate before a kick is ${\dot m_{\rm cap,bk}}\sim 9\times 10^5 (R_{\rm BH}/1~{\rm pc})^{1/2} (M_{\rm SMBH}/10^8~\Msun)^{-1/6}(m_{\rm BH}/100~{\Msun})^{-1/3}$. 
Note that the jet kinetic power can reach $L_{\rm j}\sim 10^{47}~{\rm erg/s}$ for emission from both shocked winds and AGN gas, 
by using the parameters in Eq.~\eqref{eq:l_gamma_val}, provided $r_{\rm kick}<r_{\rm trap,bk}$. 
This jet power can account for the luminosity of the breakout emission needed to reproduce the observed gamma-ray flare, assuming Doppler beaming effects. 
This finding suggests that both the observed gamma-ray and optical flares can be explained by a unified set of parameters. 
Conversely, if $r_{\rm kick}>r_{\rm trap,bk}$, unrealistically large values of $f_{\rm inc}$ may be required to produce $L_{\rm j}\sim 10^{47}~{\rm erg/s}$ (depending on ${\dot m_{\rm cap,bk}}$), which could provide a testable constraint on $f_{\rm inc}$.

We can then 
roughly assess whether the transition from the ADIOS to the ZEBRA states occurs after the kicks. With the parameters set as above: 
Before the kicks, the circularization radius is: 
$r_{\rm circ,bk}=10^{16}(R_{\rm BH}f_{\rm circ}/1~{\rm pc})[(m_{\rm BH}/M_{\rm SMBH})/10^{-7}]^{1/3}~{\rm cm}$, 
and the trapping radius is $r_{\rm trap,bk}\sim 9\times 10^{14}~{\rm cm}$ and $3\times 10^{13}~{\rm cm}$ for the shocked wind and AGN disk gas, respectively. 
In general, since $r_{\rm circ,bk}>r_{\rm trap,bk}$, 
the system remains in the ADIOS state before the kicks. 
After the kicks, 
shocks induce a  circularization radius: $r_{\rm b}\sim 7\times 10^{11}~{\rm cm}(v_{\rm kick}/500~{\rm km~s^{-1}})^{-2}$ and the trapping radius becomes 
$r_{\rm trap,ak}\gtrsim 4\times 10^{14}(m_{\rm BH}/100~\Msun)~{\rm cm}$ for $L_{\rm j}\gtrsim 10^{47}~{\rm erg~s^{-1}}$. 
This leads to $r_{\rm b}\ll r_{\rm trap,ak}$, 
which drives a transition to the ZEBRA state and enables strong jet formation after mergers.

In summary, the gamma-ray and optical counterparts can be explained with a common parameter set ($\theta_0\sim 0.05$, $m_{\rm cap,ak}\sim 10^7$, $m_{\rm BH}\sim 50$--$100~\Msun$), although 
optimistic assumptions are required to match the observed rates (\S~\ref{sec:prob_gamma}). 
The models can be ruled out if the inferred parameters are physically unrealistic--for example, $v_{\rm kick}$, $f_{\rm inc}$, $\theta_0$, and $\gamma_{\rm h,f}$ for breakout emission from CBDs, or 
$v_{\rm ej}$ and $f_{\rm inc}$ 
for cooling emission from shocked winds and AGN gas.

According to Fig.~\ref{fig:prop_l}, assuming that BHs merge in regions with low aspect ratios, 
extended long durations ($\gtrsim 10$ days) of the observed flares are consistent with the SG model. 
This favors the $\alpha$ disk model, particularly at merging locations. 
Hence, the detection of flares is useful for distinguishing between AGN models.

Optical emission may also originate from non-thermal shock emission in the AGN-disk gas \citep{Tagawa2023}. 
The observed color evolution and inter-band delays would help discriminate between these scenarios.

\subsubsection{Association probability}

\label{sec:discussions_probability}

We have shown that collisions involving jets interacting with a 
CBD, 
as well as with winds or AGN disk gas, 
can reproduce the observed characteristics of both gamma-ray and optical flares using a common set of parameters.

If the jet direction is isotropic after the merger, the probabilities of the jet colliding with winds and AGN disk gas are roughly: 
$P_{\rm col,w}\sim \Omega_{\rm w}/4\pi\sim 0.5$--0.85 \citep{Poutanen2007} and 
$P_{\rm col,AGN}\sim 0.3$ \citep{Tagawa2023}, respectively. 
Therefore, the collision probability can be estimated as $P_{\rm col}\sim 0.5$--$0.9$.

Since optical emission is likely obscured in Type~II AGNs and unobscured in Type~I AGNs, 
the detection probability of flares is reduced by the probability that the host galaxy is Type~I. We take this as: $P_{\rm typeI}\sim 0.5$ \citep[e.g.][]{Toba2021}.

Note that many flares may remain undetected if they are too faint to outshine the stochastic variability of the AGN, or if they are missed by on-going observational facilities. 
For AGNs with luminosities $L_{\rm AGN}\lesssim 10^{44}~{\rm erg~s^{-1}}$ in the SG model or $L_{\rm AGN}\lesssim 10^{45}~{\rm erg~s^{-1}}$ in the TQM model, 
the BH-driven flares are brighter than the host AGN luminosity, making most of them detectable with a maximum probability: $P_{\rm bright}\sim 1$. 
Given the uncertainties in host-AGN luminosities, we adopt $P_{\rm bright}\sim 1$.

Combining these factors, 
the probability that optical BH-driven flares are observable--under optimistic assumptions--is approximately: 
\begin{align}
P_{\rm assoc}&\sim 
P_{\rm col} P_{\rm bright} P_{\rm active} P_{\rm typeI}\nonumber\\
&\sim 0.4
\left(\frac{P_{\rm col}}{0.7}\right)
\left(\frac{P_{\rm bright}}{1}\right) 
\left(\frac{P_{\rm active}}{1}\right) 
\left(\frac{P_{\rm typeI}}{0.5}\right). 
\end{align}
Survey coverage, such as the approximately $50\%$ sky coverage by ZTF, will further reduce the detection rate by a factor of $\sim 2$.

If the duration of flares is typically shorter than the survey cadence ($\sim 3$ days for the ZTF, \citealt{Bellm_2018}), 
the probability of detection may be lower. 
Focusing on the optical flares reported by \citet{Graham2015} and \citet{Graham2023}, which have durations of $\gtrsim 10$ days, allows us to ignore this influence on the probability. 
However, it is important to note that current observations may miss a significant fraction of flares expected from AGNs with low luminosities or high angular momentum transfer (Fig.~\ref{fig:prop_l}).

Since the fraction of GW events reported to be associated with flares is lower than this estimate, 
some of the probabilities above are likely overestimated, or 
the contribution from the AGN channel to the observed GW events is small. 
These EM counterparts present intriguing targets for further exploration.

\section{Conclusions}

\label{sec:conclusions}

We explore post-merger EM flares associated with BH mergers in AGN disks. 
Our analysis focuses on 
the shock-breakout and cooling emission produced when jets (launched by BHs following mergers or through binary-single interactions) 
collide with circumbinary disks (CBDs), winds, or AGN disk gas, as well as analogous emission following SN explosions occurring in AGN disks. 

This emission can accompany mergers, as 
strong post-merger shocks in the CBD can enhance accretion rates and induce 
a transition from a long-lived ADIOS to a brief ZEBRA accretion state. 
This transition occurs when the circularization radius becomes smaller than the spherical trapping radius, 
as supported by both numerical simulations and observations. 
Additionally, this transition may resolve the overgrowth of BHs that produce bright emission, because of the very short duration of the highly-accreting post merger ZEBRA state.
The extreme hyper-Eddington accretion during this ZEBRA state enables the launch of strong jets with the aid of amplified magnetic fields resulting from the enhanced accretion. Significant emission can arise once shocks form due to collisions between jets and the surrounding gaseous media.

We applied this model to 
predict how flares driven by compact objects can be observed through current and future EM facilities. 
Furthermore, we interpreted
the optical and gamma-ray flares reported by \citet{Graham2023} and \citet{Connaughton2016}, respectively. Our main findings are summarized as follows:

\begin{enumerate}

\item 
Accretion state transitions to the ZEBRA state are associated with kicks (following mergers or binary-single interactions) for objects in AGN disks with $L_{\rm AGN}\sim 10^{43}$--$10^{45}~{\rm erg/s}$ for the SG model and $L_{\rm AGN}\gtrsim 10^{44}$ for the TQM model. 
The duty cycle of the ZEBRA state induced by these kicks 
is very small ($\sim 10^{-8}$), implying negligible net BH growth from such episodes and effectively resolving the overgrowth problem. 
Importantly, bright emission associated with BH mergers are expected to be produced from these luminous AGNs.

\item 
In less luminous AGNs, compact objects typically avoid the ZEBRA state, regardless of kicks, and grow moderately through accretion. 
In luminous AGNs ($\gtrsim 10^{45}$--$10^{46}~{\rm erg~s^{-1}}$), 
compact objects can rapidly grow into IMBHs via the ZEBRA state across a wide range of radii, even in the absence of kicks. These IMBHs then open deep gaps in AGN disks, leading to transitions to the ADIOS state. 
Such massive BHs are expected to merge with the central SMBH during quiescent phases, a process that will be constrained by future GW observatories such as LISA, TianQin, and Taiji.

\item 
Shock-cooling emission associated with jets can be bright in the optical-UV bands and its properties can help distinguish between different AGN disk models. 
The emission temperature tends to be higher, and duration shorter, 
for AGN models with more efficient angular momentum transfer, given a specific AGN luminosity. 
Emission from shocked winds and AGN disk gas, 
lasting from about an hour to a month, 
can be detected by monitoring of $\gtrsim 10^3$ AGNs with luminosities around $10^{44}$ to $10^{45}~{\rm erg~s^{-1}}$ over the course of one year, assuming BH mergers occur within these AGNs. 
From the luminosity, temperature, and duration, one can infer the size, mass, and expansion velocity of the shocked material. 
Additionally, precursor signals such as gamma-ray, X-ray, or GW emission would assist in identifying these flares.

\item 
Shock-breakout emission from the jet head can reach  
gamma-ray luminosities of $\sim 10^{44}$--$10^{47}~{\rm erg~s^{-1}}$. 
The duration and delay time of this emission are highly uncertain, ranging from $\sim 0.1$ to $10^5~{\rm s}$. This emission is detectable by MeV gamma-ray telescopes 
and may additionally occur $\sim 10^4$--$10^5~{\rm s}$ prior to a GW event caused by binary-single interactions that lead to GW capture mergers. 

\item 
At the favorable AGN luminosity of $\sim 10^{44}-10^{45}~\mathrm{erg~s^{-1}}$, cooling emission in the TQM model is hotter and shorter in duration compared to the SG model. 
Additionally, shock-breakout emission in the TQM model is dimmer across a wide range of luminosities. Unlike the SG model, the TQM model typically does not predict flares below $\sim 10^{44}~\mathrm{erg~s^{-1}}$. These trends in flare properties may provide useful diagnostics for distinguishing between AGN disk models.

\item 
Similar shock-cooling and breakout emission can also occur following SN explosions embedded in AGN disks. These are bright in the UV bands, with durations ranging from $\sim 10^3$ to $10^{5}~{\rm s}$ in the SG model. 
However, this emission is 
generally less detectable in the TQM model 
due to the smaller scale height of the disk, which results in lower cooling luminosity, 
as well as reduced gas densities that lead to fainter breakout emission. 
These signals represent promising targets for upcoming UV telescopes in thicker and denser disks.

\item 
The observed properties of candidate gamma-ray, hard X-ray, and optical counterparts to GW events can be explained within a unified model, where jets interact with CBDs, winds, or AGN disk gas using a consistent set of parameters. 
The long durations of the suggested optical flares are consistent with the AGN disk model with $\alpha$ viscosity at the merger location. 
This model offers a pathway to produce bright emission, 
which would be required to explain the flares recently claimed to be associated with GW events.

\item 
For the first time, the same model also avoids the overgrowth of the hyper-accreting BHs, due to the extreme brevity of the post-merger hyper-accreting phase.
\end{enumerate}

\vspace{\baselineskip}
In summary, we find that post-merger emission can outshine the AGN and may be detectable, and can also be useful 
for constraining BH evolution and mergers in AGN disks. 
We find that bright emission, as reported previously, can be produced without leading to issues such as overgrowth of BHs. 
Further EM observations searching for BH-driven flares potentially associated with GWs are needed across various bands. 
Additionally, targeted GRMHD simulations that clarify the transitions between accretion states, the accretion flow following mergers, jet propagation, and radiation production would be valuable for enhancing predictions.

\acknowledgments

We thank Wen-Biao Han for fruitful discussions on possible scenarios. 
H.T. is supported by 
the National Science and Technology Major Project of China (No. 2024ZD1100601) and 
the National Key R$\&$D Program of China (grant No.2024YFC2207700). 
S.S.K. was supported by 
Japan Society for the Promotion of Science (JSPS) KAKENHI 
grant Number 22K14028, 21H04487, and 23H04899, and 
the Tohoku Initiative for Fostering Global Researchers for Interdisciplinary Sciences (TI-FRIS) of MEXT's Strategic Professional Development Program for Young Researchers. 
Z.H. was supported by NASA grant 80NSSC22K0822 and NSF grant AST-2006176.

\appendix

\begin{figure*}
\begin{center}
\includegraphics[width=180mm]{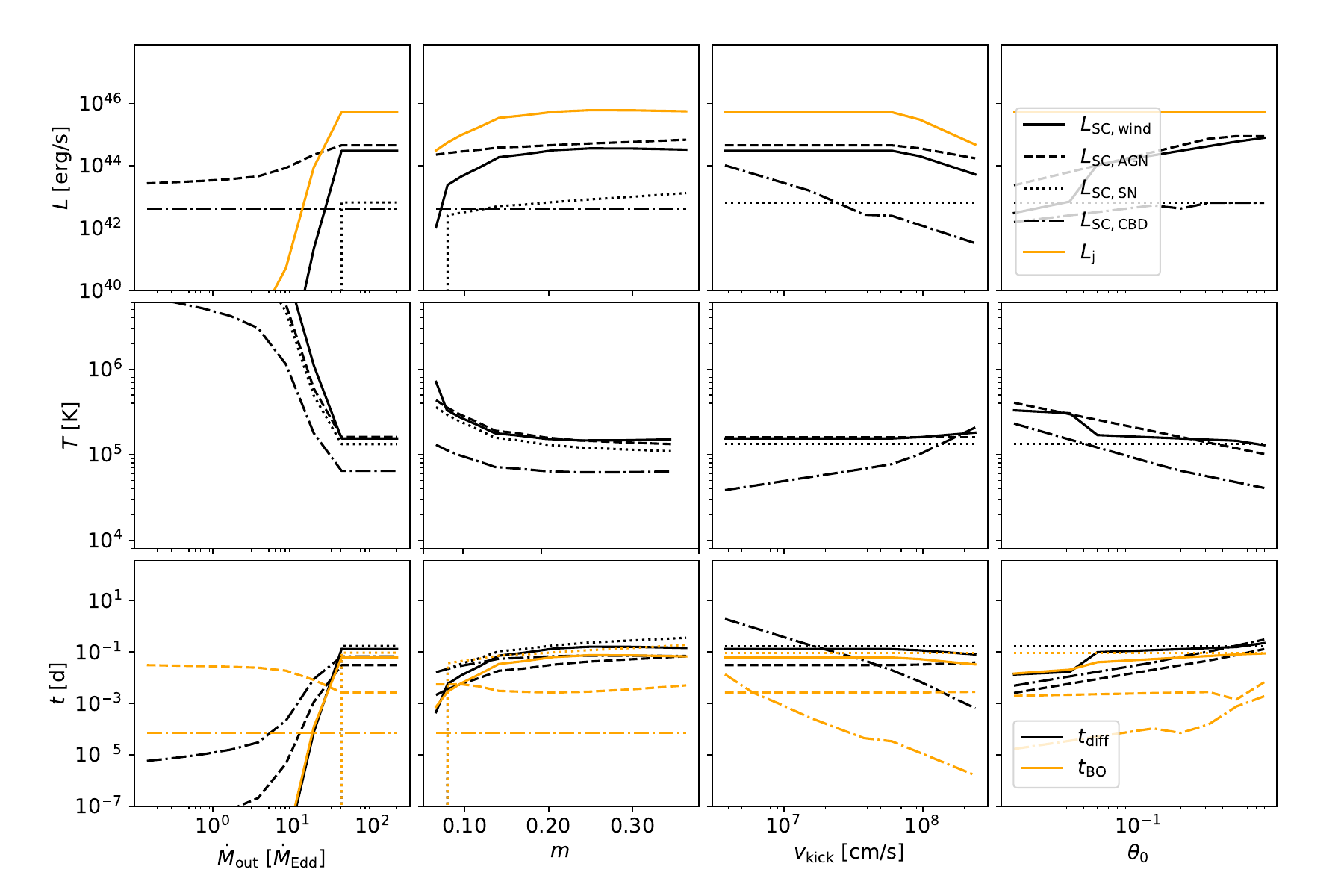}
\caption{
Same as Fig.~\ref{fig:flare_m}, but showing the dependence on several parameters. 
The left to right columns represent the dependence on 
${\dot M}_{\rm out}$, $m$, $v_{\rm kick}$, and $\theta_0$. 
}
\label{fig:flare_dep1}
\end{center}
\end{figure*}

\begin{figure*}
\begin{center}
\includegraphics[width=180mm]{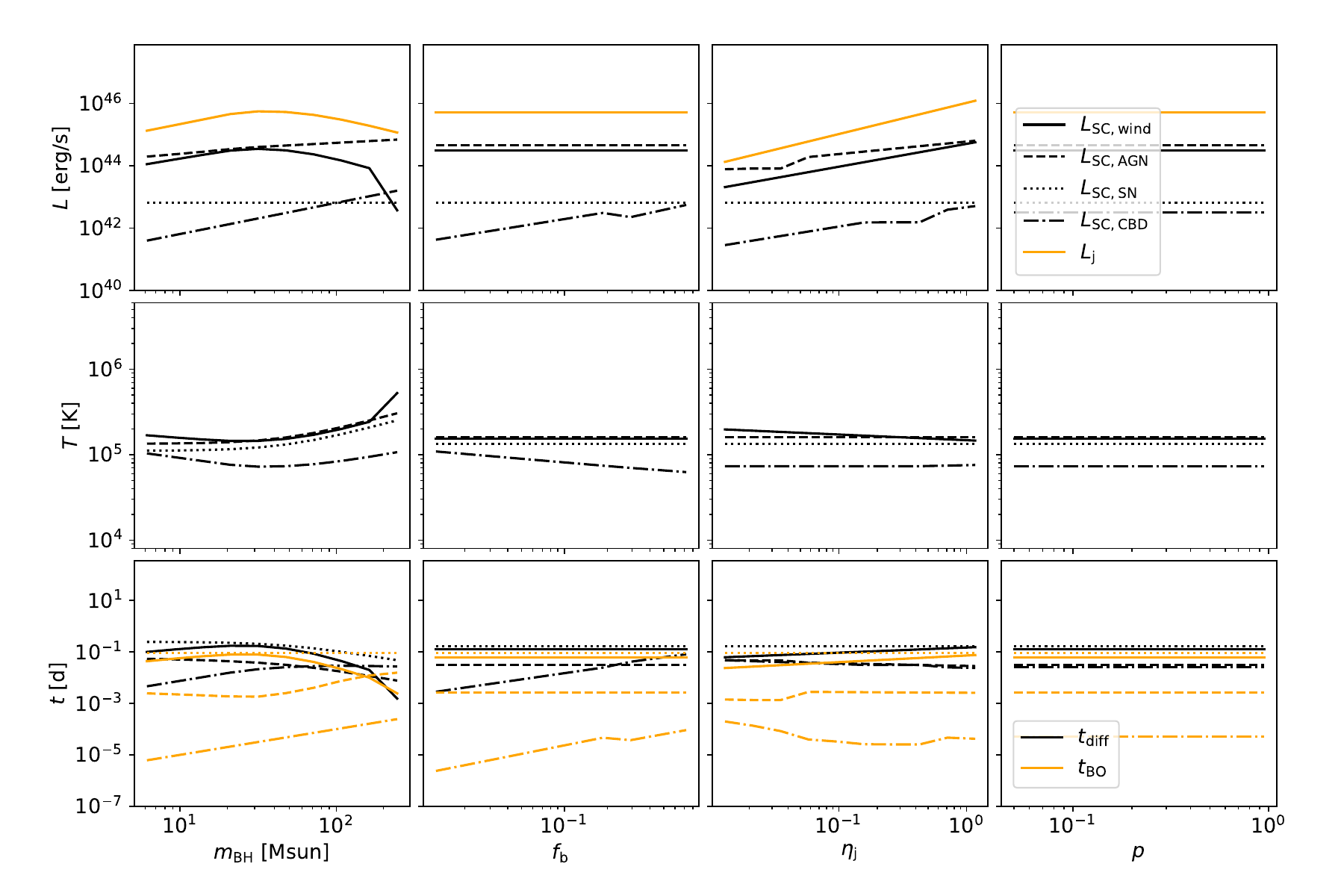}
\caption{
Same as Fig.~\ref{fig:flare_dep1}, but showing the dependence on 
$m_{\rm BH}$, $f_b$, $\eta_{\rm j}$, and $p$. 
}
\label{fig:flare_dep2}
\end{center}
\end{figure*}

\section{Parameter dependence}

\label{sec:parameter_dependence}

In this section, we present the parameter dependence of the properties of cooling emission in the fiducial settings with the TQM model 
(Figs.~\ref{fig:flare_dep1} and \ref{fig:flare_dep2}).

For high values of ${\dot M}_{\rm out}$ or $m$ cases (the first and second columns of Fig.~\ref{fig:flare_dep1}), 
shock-cooling emission from the shocked AGN disk gas 
indicate that the AGN disk becomes hot and thick due to a high heating rate from viscosity. 
This leads to an increase in luminosity and a longer diffusion timescale. 
Additionally, the high gas pressure resulting from increased heating raises the ratio $M_{\rm BO}/R_{\rm BO}$ (e.g. Eq.~C3 of \citealt{Thompson05}), which in turn reduces the radiation temperature. 
Conversely, for cases with low ${\dot M}_{\rm out}$ or $m$, gaps form due to a high aspect ratio or inefficient angular momentum transfer, rapidly decreasing the gas capture rate by the BH. 
As a result, the temperature increases and the diffusion timescale decreases for emission from all components, due to a reduction in the shocked gas mass ($m_{\rm BO}$). 

For shock-cooling emission from shocked winds, 
the high gas capture rate by the BH for high ${\dot M}_{\rm out}$ or high $m$ (the latter occurring due to gap formation) 
leads to an increase in the size of the wind as $H_{\tau}\propto {\dot M}_{\rm cap}^{3/2}$. 
This enhances 
the luminosity and duration 
while reducing the temperature. 
In terms of emission from shocked CBDs, 
the high gas capture rate for high ${\dot M}_{\rm out}$ or high $m$ also reduces the radiation temperature due to the increased mass of the CBDs.

Regarding emission induced by jets, 
As $v_{\rm kick}$ increases (the third column of Fig.~\ref{fig:flare_dep1}), 
${\dot m}_{\rm acc,ak}$ decreases (Eqs.~\ref{eq:rkick} and \ref{eq:macc_af}). 
Consequently, $L_{\rm j}$, $v_{\rm ej}$, and $L_{\rm SC}$ decrease. 
In the case of emission from shocked CBDs, 
the size decreases with increasing $v_{\rm kick}$, 
which enhances $T_{\rm SC}$ while reducing  $L_{\rm SC}$ and the timescales. 
Additionally, as $\theta_0$ decreases (the fourth column of Fig.~\ref{fig:flare_dep1}), 
both $R_{\rm BO}$ and $m_{\rm BO}$ decrease, 
resulting in lower $L_{\rm SC}$ and reduced timescales, while enhancing $T_{\rm SC}$.

As $m_{\rm BH}$ increases (the first column of Fig.~\ref{fig:flare_dep2}), 
both ${\dot m}_{\rm cap}$ and $L_{\rm j}$ increase for $m_{\rm BH}\lesssim 30~\Msun$, while decrease for $m_{\rm BH}\gtrsim 30~\Msun$ due to gap formation. 
Furthermore, as $\eta_{\rm j}$ increases (the second column of Fig.~\ref{fig:flare_dep2}), $L_{\rm j}$ also increases. 
With an increase in $L_{\rm j}$, $v_{\rm ef}$ and $L_{\rm SC}$ typically increase. 
In the case of emission from shocked CBDs, their sizes expand with increasing $m_{\rm BH}$ (Eqs.~\ref{eq:h_tau}, \ref{eq:rb}, and \ref{eq:rkick}), 
enhancing $L_{\rm SC}$. 
At high $m_{\rm BH}$, the lower ${\dot m}_{\rm cap}$ due to gap formation, 
leads to decreased timescales and an increase in $T_{\rm SC}$. 
For emission from shocked winds, $L_{\rm SC}$ is significantly reduced once a gap forms at high $m_{\rm BH}$.

With an increase in $f_{\rm b}$ (the third column in Fig.~\ref{fig:flare_dep2}), 
the size of CBDs increases (Eq.~\ref{eq:rb}), 
which enhances $L_{\rm SC}$ and the timescales while reducing $T_{\rm SC}$ for shock-cooling emission from shocked CBDs.

As $p$ increases, ${\dot m}_{\rm acc,ak}$ experiences a slight decrease, leading to a reduction in both $L_{\rm jet}$ and $L_{\rm SC}$ for the condition where $r_{\rm kick}<r_{\rm trap,bk}$. 
In the fiducial model, the properties exhibit minimal dependence on $p$ (the fourth column of Fig.~\ref{fig:flare_dep2}), 
since ${\rm min}\{1,(r_{\rm kick}/r_{\rm trap,bk})^p\} \sim 1$, given that $r_{\rm kick}\gtrsim r_{\rm trap,bk}$. 
However, the dependence on $p$ is significantly influenced by the ratio $r_{\rm kick}/r_{\rm trap,bk}$.

\section{Growth of BHs}

\label{sec:growth}

In this section, we address the overgrowth problem. 
In low luminous AGNs, 
the ADIOS state is established over a wide range of locations (\S~\ref{sec:results_accretion}, Figs~\ref{fig:ad_zeb_tqm} and \ref{fig:ad_zeb_sg}). 
If the accretion rate remains limited to around the Eddington rate with $p\sim 1$ \citep{Pan2021,Ishibashi2024,Fragile2025}, the growth of BHs in the ADIOS state during AGN phases is moderate.

Furthermore, the duration of the ZEBRA state after kicks is determined by the viscous time at $r_{\rm b}$ (see Eq.~\ref{eq:t_vis}), which is $\sim 10^5$--$10^6$ seconds in the fiducial setting. 
In the ZEBRA state, the accretion rate is $\sim 10^6$--$10^7$ times the Eddington rate (dashed black lines in Figs.~\ref{fig:radii_tqm} and \ref{fig:radii_sg}), resulting in a mass doubling time of $\gtrsim 1$--$10~{\rm yr}$. 
This implies that BHs do not experience significant growth over multiple episodes. 
Assuming BH mergers occur on a timescale of $\sim 10~{\rm Myr}$ and involve $\sim 10$ binary-single interactions before a merger \citep{Tagawa19}, 
the duty cycle of the ZEBRA state is roughly estimated to be $\sim 3\times 10^{-9}$--$3\times 10^{-8}$. 
During this period, 
the gas in the AGN disk is negligibly depleted, as the gas mass accreted onto an SMBH in $\sim 10~{\rm Myr}$ is considerably greater than the total mass of the BHs. 
Therefore, if the ADIOS state is commonly realized for most BHs in AGN disks, 
and the ZEBRA state occurs only after kicks, 
the overgrowth problem could be resolved.

In luminous AGNs, 
the ZEBRA state manifests especially in the inner and outer regions of the AGN disk, even in the absence of kicks (Figs.~\ref{fig:radii_tqm} and \ref{fig:radii_sg}), which facilitates rapid BH growth. 
One potential mechanism to reduce accretion rates is gap formation. When the disk has a short scale height, gaps can develop. 
Even if certain BHs undergo significant growth, this can lead to a decrease in the accretion rate due to gap formation resulting from that growth (Figs.~\ref{fig:mass_tqm} and \ref{fig:mass_sg}). The condition for gap formation is given by 
\begin{align}\frac{m_{\rm BH}}{M_{\rm SMBH}} \gtrsim 10^{-6}\left(\frac{h_{\rm AGN}}{0.003}\right)^{5/2} \left(\frac{\alpha}{0.1}\right)^{1/2}\end{align}\citep{Kanagawa15,Fung14}. 
Once a gap forms, the subsequent reduction in surface density can decrease the trapping radius, thereby 
encouraging BHs to transition back to the ADIOS state and limiting their growth.

Moreover, 
even if IMBHs form, as long as their formation is less efficient compared to models that permit consistent hyper-Eddington accretion and their total number remains small, this does not contradict Soltan's argument \citep{Yu2002,Tagawa2022_BHFeedback}. 
IMBHs could also merge with central SMBHs during quiescent phases due to stellar dynamical friction, with a merger timescale given by \citep{Kocsis11b}
\begin{align}\sim 10^9\left(\frac{m_\text{BH}}{1000~\Msun}\right)^{-1}\left(\frac{M_\text{SMBH}}{10^8~\Msun}\right)^{0.5}~{\rm yr}, \end{align} 
assuming a stellar density profile with a slope of -1.5 within the gravitational influence of the SMBH. Additionally, in the inner regions, IMBHs can merge with SMBHs via GW emission, with a timescale \citep{Peters64}:
\begin{align}\sim 5\times 10^7\left(\frac{R_{\rm BH}}{0.001~{\rm pc}}\right)^{4} \left(\frac{m_\text{BH}}{1000~\Msun}\right)^{-1}\left(\frac{M_\text{SMBH}}{10^8~\Msun}\right)^{-2}~{\rm yr}. \end{align}
As a result, IMBHs could merge with SMBHs within $\sim 0.1$--$1~{\rm Gyr}$, a process that can be directly constrained by future GW observatories such as the Laser Interferometer Space Antenna \citep[LISA, ][]{AmaroSeoane2017_LISA}, TianQin \citep{Luo2016_TianQin,Li2024_TianQin}, and Taiji \citep{Ruan2020_Taiji}. 
Unless a significant number of IMBHs form within the inner regions of $\lesssim 0.01~{\rm pc}$ during recent short-lived AGN episodes \citep{Su2010}, 
a scenario anticipated for SMBHs with masses around $M_{\rm SMBH}\sim 10^6~\Msun$ (Figs.~\ref{fig:mass_tqm} and \ref{fig:mass_sg}), 
the evolution of BHs in AGN disks would not contradict the dynamics of S-stars \citep{Tagawa2022_BHFeedback,Gravity2023}. 
Therefore, the overgrowth problem can likely be mitigated by considering the transitions between accretion states.

\bibliographystyle{aasjournal}
\bibliography{agn_bhm}

\end{document}